\documentclass[journal,draftclsnofoot,onecolumn,12pt,twoside]{IEEEtran}

\usepackage{graphicx}
\usepackage{amsmath, amsthm, amssymb}
\usepackage{lipsum} 
\usepackage{cite}
\usepackage{subfigure}
\usepackage{rotating}
\usepackage{fancyhdr}
\usepackage{gensymb}

\usepackage{graphicx}
\usepackage{float}
\usepackage{multicol}
\usepackage{algorithmic}
\usepackage{breqn}
\usepackage{amsmath}

\usepackage[linesnumbered,ruled]{algorithm2e}
\usepackage{color}

\usepackage{bbm}

\DeclareMathOperator{\argmin}{arg\,min}

\newtheorem{lemma}{Lemma}
\newtheorem{theorem}{Theorem}

\newtheorem{remark}{Remark}

\newtheorem{assumption}{Assumption}
\newtheorem{proposition}{Proposition}

\normalsize

\begin{document}
%
\title{Parallel Stochastic Optimization Framework for Large-Scale Non-Convex Stochastic Problems}

\author{Naeimeh~Omidvar,~\IEEEmembership{Member,~IEEE,}
        An~Liu,~\IEEEmembership{Senior Member,~IEEE,}
        Vincent~Lau,~\IEEEmembership{Fellow,~IEEE,}
        Danny~H.~K.~Tsang,~\IEEEmembership{Fellow,~IEEE,}
        and~Mohammad~Reza~Pakravan,~\IEEEmembership{Member,~IEEE}
		}


\maketitle



\begin{abstract}
In this paper, we consider the problem of stochastic optimization, where the objective function is in terms of the expectation of a (possibly non-convex) cost function 
that is parametrized by a random variable. While the convergence speed is critical for many  emerging applications, most  existing
stochastic optimization methods suffer from slow convergence. 
Furthermore, the emerging technology of parallel computing  has motivated
an increasing demand for designing new stochastic optimization  schemes that can handle parallel optimization for implementation in distributed systems.


We propose a fast parallel stochastic optimization framework that can solve a large class of possibly non-convex stochastic optimization problems that may arise in applications with multi-agent systems. In the proposed method, each agent updates its control variable in parallel, by solving a convex quadratic subproblem independently.  
The convergence of the proposed method to the optimal solution for convex problems and to a stationary point for general non-convex problems is established. 

The proposed algorithm can be applied to solve a large class of optimization problems arising in important applications from various fields, such as machine learning and wireless networks. 
%
%
%
  As a representative application of our proposed stochastic optimization framework, we focus on large-scale support vector machines and demonstrate how our algorithm can efficiently solve this problem
    , especially in modern 
    applications with huge datasets. 
    Using popular real-world datasets, we present experimental results to demonstrate the merits of our proposed framework by comparing its performance 
    to the state-of-the-art 
    in the literature. Numerical results show that the proposed method can significantly outperform the state-of-the-art methods in terms of the convergence speed while having the same or lower complexity and storage requirement.

\end{abstract}


\IEEEpeerreviewmaketitle

\section{Introduction}

Finding the minimum of a (possibly non-convex) stochastic optimization problem over a set 
in an iterative way is very popular 
in a variety of fields and applications, such as 
signal processing, wireless communications, machine learning,  social networks, economics, statistics and bioinformatics, to name just a few. 
Such stochastic problems arise 
in two different types of formulations, as will be discussed in the following. 

The first formulation type corresponds to the case where the objective function is in terms of the expectation of a cost function which is parametrised by a random (vector) variable, as shown in the following formulation:
\begin{equation}\label{eq: main stochastic optimisation_v01}
\displaystyle \min_{ \boldsymbol{x} \in \mathcal{X} } F \left( \boldsymbol{x} \right) = \mathbb{E}_{\boldsymbol{\zeta}} \left[ f \left(\boldsymbol{x}, \boldsymbol{\zeta} \right)  \right] ,
\end{equation}
where $ \boldsymbol{x} $ is the optimization variable and $  \boldsymbol{\zeta}  $ is the random variable involved in the problem. The expectation involved in the objective function may not have a closed-form expression due to the statistics of the random variables being unknown or the computational complexity of computing the exact expectation being excessively high. 
Therefore, the  deterministic optimization methods cannot solve this optimization problem. 
Such problem formulation is encountered 
 in various problems and applications in various fields such as wireless communications, signal processing, economics and bioinformatics. 
For instances of such optimization problems, you may see \cite{shuai2018stochastic, myTSPpaper, rezaei2019Journal-IoT, marti2013stochastic, movahednasab2019ICC, rezaei2018GLOBECOM, omidvar2015Globecom,  bedi2019asynchronous, omidvar2015PIMRC, movahednasab2019TCOM,  rezaei2018PIMRC,  omidvar2016cross}. 

Another category of problem formulations that utilize stochastic optimization corresponds to large-scale optimizations in which the objective function is a deterministic function, but in terms of the summation of a large number of sub-functions
, as shown in the following formulation:
\begin{equation}\label{eq: main stochastic optimisation_v02}
\displaystyle \min_{ \boldsymbol{x} \in \mathcal{X} } F \left( \boldsymbol{x} \right) = \dfrac{1}{n} \sum_{i=1}^{n} f_i \left(\boldsymbol{x} \right),    
\end{equation}
where  $ \boldsymbol{x} $ is the optimization variable, $ n $ is the number of terms in the sum and each sub-function $ f_i \left(\boldsymbol{x} \right) =  f \left(\boldsymbol{x}, \boldsymbol{y}_i \right), ~\forall i=1,\cdots,n $ is characterised by a data sample $  \boldsymbol{y}_i $.
Such 
problems naturally arise in  various applications in the machine learning area, such as classification, regression, pattern recognition, image processing, bio-informatics and social networks \cite{zhang2018energy, li2018learning, agarap2018breast, veitch2018empirical}. Moreover, these optimization problems are typically huge and need to be solved efficiently. 


Note that the above problem formulation is naturally deterministic. 
However, in large-scale problems encountered in many emerging applications dealing with big data, 
 the number of data samples $ n $ or equivalently the number of sub-functions  (and/or the size of the optimization variable $ \boldsymbol{x} $) is very large, and hence, deterministic optimization approaches (such as gradient descent \cite{boyd2009convex}) can not be utilized  to solve them, as the computational complexity of each iteration would be excessively high. 
This is because calculating the gradient of the objective function at each iteration corresponds to calculating the gradients of all of these sub-functions, which is of huge complexity and may not be practical. 
%
As such, the extremely large size of datasets is a critical challenge for solving large-scale problems arising in emerging applications, which makes
 the classical optimization methods intractable. Consequently, 
there is an increasing need 
to develop new methods with tractable complexity 
to handle the large-scale problems efficiently\cite{BigData2014convex}.

To address the aforementioned challenge, 
 stochastic optimization approaches are used. Under such approaches, at each iteration, instead of calculating the \textit{exact gradient} of the objective function, a \textit{stochastic gradient}  will be calculated by randomly picking one sub-function (corresponding to one randomly chosen or arrived data sample) and calculating the gradient of that sub-function as a stochastic approximation of the true gradient of the objective function. As such, the computational complexity of gradient calculation is $ 1/n^{\mathrm{th}} $ that of deterministic gradient methods, and is independent of the number of data samples.  
Moreover, note that the intuition behind utilising stochastic optimization 
methods instead of deterministic 
approaches for 
the deterministic problem formulation in \eqref{eq: main stochastic optimisation_v02} is that 
the objective function in \eqref{eq: main stochastic optimisation_v02} can be viewed as the sample average approximation of the stochastic objective function 
described in \eqref{eq: main stochastic optimisation_v01}, where each sample function $ f_i \left(\boldsymbol{x} \right) $ corresponds to a realisation of the random cost function $ f \left(\boldsymbol{x}, \boldsymbol{\zeta} \right)  $ (i.e., under a fixed realisation $ \boldsymbol{\zeta}=\boldsymbol{y}_i $) 
\cite{BigData2014convex}. 

{\color{black}{Having encountered the above two forms of
 the optimization problems (that require stochastic optimization methods) in different areas,}} researchers from various communities, including but not limited to optimization and pure mathematics
, neural networks
, and machine learning fields
, have been working on stochastic optimization algorithms. Specifically, they aim to propose new 
 stochastic optimization algorithms that meet the new requirements of emerging and future applications
, as will be discussed in the following.

First of all, the main requirement for designing new stochastic optimization algorithms  
is \textit{fast convergence}. This is because the new stochastic optimization problems encountered in emerging and future applications are mainly large-scale problems that need to be handled in a very short time; otherwise, their obtained solutions may become out-dated and hence, no longer valid  at the time of convergence. Therefore,  efficient algorithms that can converge very fast are essential. 


Furthermore, in many new applications, there are 
multiple agents that have control over different variables 
and need to optimise the whole-system performance together in a distributed way. 
  These applications and scenarios deal with distributed systems in which the associated stochastic optimization problems need to be iteratively solved by a parallel stochastic optimization method.{\footnote{Such applications arise in various fields, for example, multi-agent resource allocation in wireless interference systems \cite{Daniel2016parallel}, peer-to-peer networks, adhoc networks  and cognitive radio systems  in wireless networks, parallel-computing data centres in big data, and large-scale distributed systems in economics and statistics, to name just a few.}} 
In addition, with the recent advances in  multi-core parallel processing technology, 
it is increasingly desirable to develop parallel algorithms 
that allow various control variables to be simultaneously updated at their associated agents,  at each iteration.

The 
above requirements have not been truly 
addressed by the classical stochastic optimization algorithms. Therefore,  new algorithms that can efficiently solve large-scale stochastic optimization problems with fast convergence 
and support parallel optimization of the involved control variables are needed. The existing algorithms that aim to meet such requirements 
can be 
categorised into three 
groups: stochastic gradient-based, stochastic majorization-minimization and stochastic parallel decomposition methods, as discussed in the following sections.


\subsection{Stochastic Gradient-Based Algorithms}

Stochastic gradient-based methods are variations of the well-known  stochastic gradient descent (SGD) method 
\cite{Shamir12, Shamir2013}.  In the dynamic equations of these methods, instead of using the exact gradient of the objective function, which is not available in problem formulations in the general form of \eqref{eq: main stochastic optimisation_v01} or is not practical to calculate in  problem formulations in the general form of \eqref{eq: main stochastic optimisation_v02}, a noisy estimate of the gradient is used. Such a noisy gradient estimate, which is referred to as a \textit{stochastic gradient}, 
is usually obtained by calculating the gradient of the observed sample function $ f \left(\boldsymbol{x}, \boldsymbol{\zeta} \right)  $ for the observed realisation of the random variable $ \boldsymbol{\zeta} $ in the case of the problem formulation in \eqref{eq: main stochastic optimisation_v01}, or the gradient of the sub-function $ f_i \left(\boldsymbol{x} \right) $ for the randomly chosen index $ i $ in the case of the problem formulation in \eqref{eq: main stochastic optimisation_v01}.  As the negative of an stochastic gradient is used as the update direction at each iteration, these methods are called stochastic gradient-based.

Various numerical results show that using 
such update direction suffers from slow convergence. This is mainly because the update direction may not be a tight approximation of the true gradient, and hence converges to the true gradient slowly.
To improve the slow convergence of the stochastic gradient-based update direction,  acceleration techniques have been proposed \cite{polyak1992acceleration}, including 
averaging over the iterates \cite{Shamir12}, $ \alpha$-suffix averaging \cite{Shamir12}, and polynomial-decay averaging \cite{Shamir2013}. However, although some of these methods have been proved to achieve the optimal convergence rate for a considered class of problems \cite{Shamir12}, their obtained rate is  asymptotic, and numerical experiments on a large number of problems show that in practice, their improvement is not significant compared to the SGD method
\cite{Shamir2013}. 
%
%
%
Furthermore, the convergence of the aforementioned works is proved mainly for the special case of strongly convex objective functions and may not work well for other convex functions or the general class of non-convex problems. 
The existing works that also consider non-convex 
 problems are mainly designed for the unconstrained case only \cite{bertsekas2000gradient,tsitsiklis1986distributed}. In fact, under some strict requirements on step-sizes \cite{nemirovski2009robust}, \cite{Shamir12}, these works analyze a descent convergence of the algorithm for the unconstrained case. However, such analysis may not be valid in the presence of the projection which is 
involved due to the existence of the constraints. 

Finally, another popular acceleration method to speed up the convergence, especially in training algorithms of deep networks such as ADAM algorithm \cite{ADAM_2014}, is mini-batch stochastic gradient estimation. 
However, with the rapid growth of data and increasing model complexity, it still shows a slow convergence rate, while the per-iteration complexity of mini-batch algorithms grows linearly with the size of the batch \cite{accelerating_minibatch_SGD_2019}. 

In this paper, we are interested in the general class of non-convex stochastic optimization problems and aim to propose a fast converging stochastic optimization algorithm that can handle 
constrained optimization problems, as well. 

\subsection{Stochastic Majorization-Minimization }



 
{\color{black}{To better support non-convex stochastic optimization problems, stochastic majorization-minimization method 
is also widely used.}}
This method is basically a non-trivial extension of majorization-minimization (MM) method for the deterministic problems. 
MM is an optimization method to minimise a \textit{possibly non-convex} function by iteratively minimising a convex upper-bound function that serves as a surrogate for the objective function \cite{lange2000optimization}. The intuition behind the MM method is that minimising the upper-bound surrogate functions over the iterations monotonically drives the objective function value down. Because of the simplicity of the idea behind MM, it has been  popular for a wide range of problems in signal processing and machine learning applications, and there are many existing algorithms that use this idea, including the expectation-maximisation \cite{cappe2009line}, DC programming \cite{gasso2009recovering}
 and proximal algorithms \cite{wright2009sparse,beck2009fast}. 
 However, extension of MM approaches to the stochastic optimization case is not trivial. Because in a stochastic optimization problem, there is no closed-form expression for the objective function (due to the expectation involved), and hence, it is difficult 
to find the required upper-bound convex approximation (i.e., the surrogate function) for the objective function. 

 To address the aforementioned issue, stochastic MM \cite{mairal2013stochastic}, a.k.a., stochastic successive upper-bound minimization (SSUM) \cite{razaviyayn2016stochastic},
 is proposed as a new class of algorithms for large-scale non-convex optimization problems, which extends the idea of MM to stochastic optimization problems, in the following way: 
 Instead of for the original objective function, stochastic MM tries to find a majorizing surrogate 
 for the observed sample function 
 at each iteration. 
 The currently and previously found instance surrogate functions (a.k.a. sample surrogate functions) are then incrementally combined together to form an approximate surrogate function for the objective function, which will then be minimised to update the iterate. 

Under some conditions, mainly on the instance surrogate functions, 
it has been shown that the approximate surrogate function evolves to be a majorizing surrogate for the expected cost in the objective function of the original stochastic optimization problem \cite{mairal2013stochastic}. The major condition is that the instance surrogate function should be an upper-bound for the observed sample cost function at each iteration \cite{razaviyayn2016stochastic}. 
In some of the related works in this area, this condition needs to be satisfied locally, while in most of the existing works, they require the surrogate function to be a global upper-bound \cite{razaviyayn2016stochastic}.  


 It should be noted that although the upper-bound requirement is fundamental for the convergence of these methods, 
 it is restrictive in practice. This is mainly because finding a proper upper-bound 
 for the sample cost function at each iteration may be difficult itself. 
Therefore, although this upper-bound can facilitate the minimization at each iteration, in general finding this upper-bound will increase the per-iteration computational complexity of the algorithm. Consequently, 
  minimising the approximate surrogate function at each iteration is only practical when these surrogate functions are simple enough to be easily optimized
, e.g., when they can be parametrised with a small number of variables \cite{razaviyayn2016stochastic}. Otherwise, the complexity of stochastic MM may dominate the simplicity of the idea behind it, making the method impractical.

In addition to the aforementioned complexity issue, 
stochastic MM methods may not be implementable in parallel. As in general, it is not guaranteed that the resulting problem of optimising the approximate surrogate function at each iteration is decomposable with respect to 
 the control variables of different agents, a centralised implementation might be required. Consequently, this method may not be 
 suitable for distributed systems and applications that need parallel implementations.

Motivated to 
address these issues of stochastic MM, in this paper, we propose a stochastic scheme with an approximate surrogate that not only is easy to be obtained and minimised at each iteration, but also can be easily decomposed in parallel. 
Specifically, the instance surrogate function in our method 
does not have to be an upper-bound. Moreover, it can be easily calculated and optimized at each iteration with low complexity, as will be seen in the next section. 
However, it brings new challenges that need to be tackled in order to prove the convergence of the proposed method. Most importantly, since the instance surrogate function in our algorithm is no longer an upper-bound, unlike in stochastic MM, the monotonically decreasing property cannot be utilized in our case. Therefore, it is more challenging 
to show that the approximate surrogate function will eventually become an upper-bound for the expected cost in the objective function. We will show how we tackle these challenges in later sections.

\subsection{Parallel Stochastic Optimization}

As explained before, there is an increasing need to design new stochastic optimization algorithms that enable parallel optimization of the control variables by different agents 
in a distributed manner. Note that many gradient-based methods are parallelisable in nature \cite{Shamir12,Shamir2013, nesterov2013gradient, necoara2013efficient, tseng2009coordinate}. 
However, as mentioned before, they suffer from low convergence speed in practice \cite{razaviyayn2014parallel, facchinei2014flexible}. 
%

There are quite few works on parallel stochastic optimization in the literature. The authors in  \cite{Daniel2016parallel} have proposed a stochastic parallel optimization 
method that decomposes the original stochastic (non-convex) optimization problem into parallel deterministic convex sub-problems, where the sub-problems are then solved independently by different agents in a distributed fashion. 
%
%
%
%
The proposed method here 
differs from that in \cite{Daniel2016parallel} in the following ways. 
Firstly, unlike the proposed method, the algorithm in \cite{Daniel2016parallel} requires a weighted averaging of the iterates, which slows down the convergence, in practice. 
Secondly, in our proposed method, the 
objective 
function  approximated with an incremental convex approximation that is easier to calculate and minimize, with less computational complexity than that in \cite{Daniel2016parallel}, for an arbitrary objective function in general. 
The aforementioned differences will be elaborated more, later in Section \ref{sec: comparison_with_works}.

\subsection{Contributions of This Paper} 

In this paper, we address the 
aforementioned issues of the existing works and propose a fast converging and low-complexity parallel stochastic optimization 
algorithm for general non-convex stochastic optimization problems. 
The main contributions of this work can be summarised as follows.
\begin{itemize}
\item \textbf{A stochastic convex approximation method for general (possibly non-convex) stochastic optimization problems with guaranteed convergence
:} We propose a stochastic convex approximation framework that can solve general stochastic optimization problems (i.e., without the requirement to be strongly convex or even convex), with low complexity. We 
 analyze the convergence of the proposed framework  for both cases of convex and non-convex stochastic optimization problems, and prove its convergence to the optimal solution for convex problems and to a stationary
point for general non-convex problems.

\item \textbf{A general framework for parallel stochastic optimization}: 
We show that our proposed method can be applied for parallel decomposition of stochastic multi-agent optimization problems arising in distributed systems. Under our proposed method, the original (possibly non-convex) stochastic   optimization problem is decomposed into parallel deterministic convex sub-problems and each sub-problem is then solved by the associated agent, in a distributed fashion. Such a parallel optimization approach can be highly beneficial for reducing computational complexity, especially in large-scale optimization problems.


\item \textbf{Applications to solve large-scale stochastic optimization problems with fast convergence:} We show by simulations that our proposed framework can efficiently solve large-scale stochastic optimization problems 
in the area of machine learning.
Comparing the proposed method to the state-of-the-art methods 
 shows that ours significantly outperforms the state-of-the-art methods in terms of convergence speed, while maintaining low computational complexity and storage requirements. 


\end{itemize} 

The rest of the paper is organised as follows. Section \ref{sec: ProbFormulation} formulates the problem. 
Section \ref{sec: ProposedMethod} introduces the proposed parallel stochastic optimization 
framework and presents its convergence results for both the convex and non-convex cases. 
In Section \ref{sec:SVM_Simulations}, we illustrate an important application example of the proposed framework in the area of machine learning, and show how the proposed method can efficiently solve  the problem in this application with low complexity and high convergence speed.
Simulation results and comparison to the state-of-the-art methods are presented in Section \ref{sec: SVM_sim}. Finally, Section 
\ref{sec: SVM_conclusion} 
concludes the chapter.

\section{Problem Formulation}\label{sec: ProbFormulation}


Consider a multi-agent system composed of $ I $ users, each independently controlling a strategy vector $ \boldsymbol{x}_l \in \mathcal{X}_l, ~l=1,\cdots,I $, that aim at solve the following stochastic optimization problem together in a distributed manner:
\begin{equation}\label{eq: main stochastic optimisation}
\displaystyle \min_{ \boldsymbol{x} \in \mathcal{X} } F \left( \boldsymbol{x} \right) \triangleq \mathbb{E}_{\boldsymbol{\zeta}} \left[ f \left(\boldsymbol{x}, \boldsymbol{\zeta} \right)  \right] ,
\end{equation}
where $  \boldsymbol{x} \triangleq  \left(  \boldsymbol{x}_l \right)_{l=1}^I  $ is the joint strategy vector and $ \mathcal{X} \triangleq \mathcal{X}_1 \times \cdots \times \mathcal{X}_I $ is the joint strategy set of all users. Moreover, the objective function $ f \left( \boldsymbol{x} \right) $ is the expectation of a 
sample cost function $  f \left(\boldsymbol{x}, \boldsymbol{\zeta} \right) : \mathcal{X} \times \Omega \rightarrow \mathbb{R} $ which depends on the joint strategy vector $ \boldsymbol{x} $ and a random vector $ \boldsymbol{\zeta}  $.  The random vector $ \boldsymbol{\zeta}  $ is defined on a probability space denoted by $ \left( \Omega, \mathcal{F}, P \right) $, where the probability measure $ P $ is unknown. 


Such problem formulation is very general and includes many optimization problems as its special cases. Specifically, since the objective function is not assumed to be convex over the set $ \mathcal{X} $, the considered optimization problem can be non-convex in general. 
The following assumptions are made throughout this paper, which are classical in the stochastic optimization literature and are satisfied for a large class of problems \cite{nemirovski2009robust}.



\begin{assumption}[Problem formualtion structure]\label{assump: stoch opt formulation}
For the  problem formulation in \eqref{eq: main stochastic optimisation}, we assume
\begin{itemize}
\item[a)] The feasible sets $ \mathcal{X}_l, ~ \forall l=1,\ldots,I $ are compact and convex;\footnote{This assumption guarantees that there exists a solution to the considered optimization problem.
}
\item[b)] For any random vector realisation $  \boldsymbol{\zeta} $, the  
sample cost function $  f \left(\boldsymbol{x}, \boldsymbol{\zeta} \right) $ is 
continuously differentiable over $ \mathcal{X} $ and has a Lipschitz continuous gradient;
\item[c)] The objective function $ F \left( \boldsymbol{x} \right)  $ has a Lipschitz continuous gradient with constant $ L_{\nabla F} < +\infty $.

\end{itemize}
\end{assumption}

We aim at designing a distributed iterative algorithm with low complexity, fast convergence and low storage requirements that solves the considered stochastic optimization problem when the distribution of the random variable $ \boldsymbol{\zeta} $ is not known or it is practically impossible to accurately compute the expected value in \eqref{eq: main stochastic optimisation} and hence, we only have access to the observed samples (i.e., realisations) of the random variable $ \boldsymbol{\zeta} $.

\section{Proposed Parallel Stochastic Optimization Algorithm}\label{sec: ProposedMethod}

\subsection{Algorithm Description}

Note that the considered problem in \eqref{eq: main stochastic optimisation} is possibly non-convex. Moreover, due to the expectation involved, its objective function does not have a closed-form expression.
These two issues make finding the stationary point(s) of this problem very challenging. 
In the following, we tackle these challenges and propose an iterative decomposition algorithm that solves the problem in a distributed manner, where the agents update and optimise their associated surrogate functions in parallel. In this way, the original objective function is replaced with some incremental strongly convex surrogate functions which  will then be updated and optimized by the agents in parallel. 
Note that the proposed method can also 
support the mini-batch case. Therefore, for the sake of generality, we consider the 
mini-batch stochastic gradient estimation, where the size of the mini-batch, parametrized by $ B \in \{ 1, 2, \ldots \} $, can be chosen to achieve a good trade-off between the per-iteration complexity and the convergence speed. 

\begin{algorithm}
  \begin{algorithmic}[1]

    \STATE {\textbf{Input:} $ \boldsymbol{x}^0 \in \mathcal{X} $
    ; $ \left\lbrace \omega_k \right\rbrace_{k\geq 1} $
    ;  $ \left\lbrace \alpha_k \right\rbrace_{k\geq 1} $
	; $ k=0 $; $ \boldsymbol{h}^0_l=\boldsymbol{0}, ~ \forall l=1,\ldots,I $; mini-batch size $ B \geq 1 $.}
    
    \STATE If $ \boldsymbol{x}^k $ satisfies a suitable termination criterion: STOP. \label{new_iteration} 
    
    
  \STATE 
  Observe the 
  batch of independent random vector realisations $ \boldsymbol{Z}^k = \{ \boldsymbol{\zeta}^k_b \}_{b=1}^{B} $, 
  and hence, its 
  associated 
  sample function 
  $ f^k \left( \boldsymbol{x} \right) = \frac{1}{B} \sum_{b=1}^B f \left(\boldsymbol{x}, \boldsymbol{\zeta}^k_b \right) $.

      

    \FOR {$l = 1$ to $I$}
            
            \STATE Update 
            the vector 
      \begin{equation}\label{eq: h^k update}
      \boldsymbol{h}_l^k =  \left( 1-\omega_k \right) \boldsymbol{h}_l^{k-1} + \omega_k \nabla_l f^k \left(\boldsymbol{x}^{k-1} \right).
      \end{equation}       
         %
            %
            \STATE Define the surrogate function 
                        \begin{equation}\label{eq: hat_f_k update}
                        \hat{f}_l^k(\boldsymbol{x}_l) = \dfrac{1}{2 \alpha_k} \parallel \boldsymbol{x}_l - \boldsymbol{x}_l^{k-1}  \parallel^2 + \langle \boldsymbol{h}^k_l , \boldsymbol{x}_l - \boldsymbol{x}_l^{k-1}  \rangle.
                        \end{equation}
                        
                        \STATE Compute 
                        \begin{equation}\label{eq: x^k update}
                        \boldsymbol{x}_l^k =  \argmin_{\boldsymbol{x}_l \in \mathcal{X}_l} \hat{f}_l^k(\boldsymbol{x}_l).
                        \end{equation}
                    \ENDFOR
                  
                  \STATE $ k \rightarrow k+1$, and go to  line \ref{new_iteration}.
            %
            \end{algorithmic}
                \caption{Stochastic Parallel Decomposition Algorithm}
                    \label{alg: main}
            \end{algorithm}

The proposed parallel decomposition method is described in Algorithm \ref{alg: main}. The iterative algorithm proceeds as follows: As each iteration $ k $,  a batch of size $ B $ of random vectors 
$  \boldsymbol{\zeta}^k_b , ~ \forall b=1,\ldots,B $ 
is realised, and accordingly, each agent $ l, ~ \forall l=1,\ldots, I $ calculates the gradient of the 
the associated sample function 
$ f^k \left(\boldsymbol{x} \right) $ with respect to its control variable $ \boldsymbol{x}_l $ 
at the latest iterate (i.e.,  $ \nabla_l f^k \left( \boldsymbol{x}^{k-1} \right) $). Then, 
using its newly calculated gradient and the previous ones, agent $ l $ incrementally updates a vector $ h_l^k $ which is an estimation of the exact gradient of the original objective function with respect to $ \boldsymbol{x}_l $. It will be shown that this vector eventually converges to the true gradient. Using this gradient estimation and the latest iterate (i.e., $ \boldsymbol{x}_l^{k-1} $), agent $ l $ then constructs 
a quadratic deterministic surrogate function, as in \eqref{eq: hat_f_k update}. 
The surrogate function is then minimised by agent $ l $ to update the control variable $ \boldsymbol{x}^k_l  $. 
In this way, through solving 
deterministic quadratic sub-problems  in parallel, each user minimises a sample convex approximation of the original non-convex stochastic function. 

Note that the first term in the surrogate function \eqref{eq: hat_f_k update} is the proximal regularisation term and makes the surrogate function strongly convex, with parameter $ \dfrac{1}{\alpha_k} $. Moreover, the role of the second term is to incrementally estimate the unavailable exact gradient of the original objective function at each iteration (i.e., $ \nabla_{\boldsymbol{x}} F \left(\boldsymbol{x}^{k-1} \right) $), using the sample gradients collected over the iterates so far. 

According to \eqref{eq: h^k update}, the direction $ \boldsymbol{h}^k $ is recursively updated based on the previous direction $ \boldsymbol{h}^k $ and the newly observed stochastic gradient $ \nabla f^k \left(\boldsymbol{x}^{k-1} \right) $.
%
Under proper choices of the sequences  $ \left\lbrace \omega_k \right\rbrace_{k=1}^\infty $ and $ \left\lbrace \alpha_k \right\rbrace_{k=1}^\infty $, it is expected that 
such incremental estimation of the exact gradient 
becomes more and more accurate as $ k $ increases, and gradually, it will converge to the exact gradient (This will be shown later on by Lemma \ref{lem: h_k  --> nabla f}).

\subsection{Convergence Analysis of the Proposed Method}

In the following, we present the main convergence results of the proposed Algorithm \ref{alg: main}. Prior to that, let us state the following assumptions on the noise of the stochastic gradients as well as the sequences $ \left\lbrace \omega_k \right\rbrace $ and $ \left\lbrace \alpha_k \right\rbrace $. 

\begin{assumption}[Unbiased gradient estimation with bounded variance]\label{assump: noise}
For any iteration $ k \geq 0 $, the following 
results hold almost surely: 
$ 1\leq \forall b \leq B $,
\begin{itemize}
\item[a)] $ \mathbb{E} \left[ \nabla F \left(\boldsymbol{x}^{k} \right) - \nabla f \left(\boldsymbol{x}^{k}, \boldsymbol{\zeta}^{k}_b \right) \left| \mathcal{F}_k \right. \right] = \boldsymbol{0} $,
\item[b)] $ \mathbb{E} \left[ \parallel \nabla F \left(\boldsymbol{x}^{k} \right) - \nabla f \left(\boldsymbol{x}^{k}, \boldsymbol{\zeta}^{k}_b \right) \parallel^2 \left| \mathcal{F}_k \right. \right] < \infty $,
\end{itemize}
where $ \mathcal{F}_k \triangleq \left\lbrace  \boldsymbol{x}^{0}, \boldsymbol{Z}^{0}, \ldots,  \boldsymbol{Z}^{k-1}  \right\rbrace $ denotes the past history of the algorithm up to iteration $ k $.
\end{assumption}

Assumption \ref{assump: noise}-(a) indicates that the instantaneous gradient is an unbiased estimation of the exact gradient at each point, and Assumption \ref{assump: noise}-(b) indicates that the variance of such noisy gradient estimation is bounded. It is noted that these assumptions are standard and very common in the literature for instantaneous gradient errors \cite{ram2009incremental,zhang2008impact}. 
Moreover, it can be easily 
verified that if the random variables $ \boldsymbol{\zeta}^{k}_b, ~\forall k \geq 0,  1 \leq \forall b \leq B  $  are bounded and identically distributed, then these assumptions are automatically satisfied \cite{nemirovski2009robust}. Finally, Assumption  \ref{assump: noise}  clearly implies  that the gradient of the observed sample function $ f^k \left(\boldsymbol{x} \right) $ at the current iterate $ \boldsymbol{x}^k $ is an unbiased estimation of the true gradient of the original objective function (However, note that $ \boldsymbol{h}^k $ is not an unbiased estimation for finite $ k $), with a finite variance. 

\begin{assumption}[Step-size sequences constraints]\label{assump: step-sizes}
The sequences $ \left\lbrace \omega_k \right\rbrace $ and $ \left\lbrace \alpha_k \right\rbrace $ 
satisfy the following conditions:
\begin{itemize}
\item[a)] $ \omega_1=1 $, $ \sum_{k=1}^\infty \omega_k = \infty $ and $ \sum_{k=1}^\infty {\omega_k}^2 < \infty $,
\item[b)] $ \sum_{k=1}^\infty \alpha_k = \infty $, $ \sum_{k=1}^\infty {\alpha_k}^2 < \infty $ and $ \alpha_k/\omega_k \rightarrow 0 $.
\end{itemize}
\end{assumption}

The following theorem states a preliminary convergence result of the proposed Algorithm \ref{alg: main} for the general (possibly non-convex) stochastic optimization problems.

\begin{theorem}
\label{th: conv_nonconvex}
For the general (possibly non-convex) objective function $ F \left(.\right) $ and under Assumptions \ref{assump: stoch opt formulation}--\ref{assump: step-sizes}, there exists a 
subsequence $ \left\lbrace \boldsymbol{x}^{k_j} \right\rbrace $ of the iterates generated by Algorithm \ref{alg: main} that converges to a stationary point of Problem \eqref{eq: main stochastic optimisation}.		
%
\end{theorem}

The following theorem shows the convergence of the proposed algorithm to the optimal solution for the case of \textit{convex} stochastic optimization problems.

\begin{theorem}{\textbf{(Convex Case)}}\label{th: conv_convex}
For convex objective function $ F \left(.\right) $ and under Assumptions \ref{assump: stoch opt formulation}-\ref{assump: step-sizes}, every limit point of the sequence $ \left\lbrace \boldsymbol{x}^k \right\rbrace $ generated by Algorithm \ref{alg: main} (at least one limit point exists) 
converges to an optimal solution of Problem \eqref{eq: main stochastic optimisation}, denoted by $ \boldsymbol{x}^\ast $,  almost surely, i.e., 
\begin{equation}
\displaystyle\lim_{k \rightarrow \infty} F \left( \boldsymbol{x}^k \right) - F \left( \boldsymbol{x}^\ast \right) =0.
\end{equation}
\end{theorem}

The last theorem establishes the convergence of the proposed algorithm to a stationary point for the general case of \textit{non-convex} stochastic optimization problems.

\begin{theorem}{\textbf{(Non-convex case)}}\label{th: conv2_nonconvex}
For the general (possibly non-convex) objective function $ F \left(.\right) $, 
if there exists a subsequence of the iterates generated by Algorithm \ref{alg: main} which converges to a strictly local minimum,
 then the sequence $ \left\lbrace \boldsymbol{x}^k \right\rbrace $ generated by Algorithm \ref{alg: main} converges to that stationary point, almost surely.
\end{theorem}

\subsection{Comparison with the Existing Methods} \label{sec: comparison_with_works}



It should be noted that unlike in SGD, the  
stochastic gradient 
vector $ \boldsymbol{h}^k $ in our method is not an unbiased estimation of the exact gradient, for finite $ k $. This fact adds to the non-trivial challenges in proving the convergence of the proposed algorithm since the expectation of the update direction is no longer a descent direction. 
To tackle this challenge, 
we prove that this biased estimation 
asymptotically converges to 
the exact gradient, with probability one. 

Moreover, unlike the stochastic MM methods discussed before, the considered surrogate functions are not necessarily an upper-bound for the observed sample function, but will eventually converge to be a global upper-bound for the expectation of the sample functions. 
In addition, the considered surrogate functions can be computed and optimized with low complexity for any optimization problem of the form in \eqref{eq: main stochastic optimisation}. These advantages address the complexity issues of 
the stochastic MM methods 
that discussed before. 

Furthermore, the proposed method here 
differs from that in \cite{Daniel2016parallel} in the following ways: 
Firstly, the algorithm proposed in \cite{Daniel2016parallel} requires weighted averaging of the iterates, where the associated vanishing factor 
is assumed to be much faster diminishing 
than the vanishing factor involved in the incremental estimate of the exact 
gradient of the objective function. 
In fact, averaging over the iterates helps to average out the noise involved in the gradient estimation  of the stochastic objective function, and hence, is used as a crucial step for proving the convergence of the method proposed in 
\cite{Daniel2016parallel}. 
However, in practice, such averaging over the iterates makes the convergence of the algorithm slower, as the weight for the approximate solution found at the current iteration converges  to zero very fast. 
 Therefore, the effect of a new approximation point found by solving the approximate surrogate function at the current iteration would very quickly become negligible. 
%
%
%
Such a step, which is fundamental for the convergence of the proposed scheme in \cite{Daniel2016parallel}, is no longer used in the proposed algorithm. Although this makes the convergence proof of our proposed method challenging, it contributes to the significantly faster convergence of the proposed scheme compared to the scheme in \cite{Daniel2016parallel}, as can be verified by the numerical results in Section \ref{sec: SVM_sim}.
%

Secondly, the surrogate function at each agent is obtained from the original objective function by replacing the convex part of the expected value with its incremental sample function 
and the non-convex part with a convex local estimation. However, in our proposed method, the expectation of the whole objective function is approximated with an 
incremental convex approximation, which can be easily calculated and optimized with low complexity, at each iteration. This contributes to the lower complexity of the proposed scheme here compared to the one in \cite{Daniel2016parallel}, for an arbitrary objective function in general. 

\section{An Application Example of the Proposed Method in Solving Large-Scale Machine Learning Problems}
\label{sec:SVM_Simulations} 

 Optimization is believed to be one of the important pillars of machine learning \cite{bottou2018optimization}. This is mainly due to the nature of machine learning systems, where a set of parameters need to be optimized based on currently available data so that the learning system can make decisions for yet unseen data. 
 Nowadays, many challenging machine learning problems in a wide range of applications are relying on optimization methods, and designing new methods that are more widely applicable to modern machine learning applications is highly desirable. 
One of the important problems which appear in many 
machine learning applications with huge datasets is large-scale support vector machines (SVMs). In this section, we demonstrate how our proposed optimization algorithm can efficiently solve this problem. We compare the performance of the proposed algorithm to the state-of-the-art methods in the literature and present experimental results to demonstrate the merits of our proposed framework. 
SVMs are 
one of the most prominent machine learning techniques for a wide range of 
classification problems in  machine learning applications, such as cancer diagnosis in bioinformatics, image classification, face recognition in computer vision and text categorisation in document processing \cite{text_categoration_SVM,Daniel2015}. The SVM classifier is formulated by the following 
 optimization problem{\footnote{Note that although SVM problem formulation is widely used in the literature to show the performance of the optimization algorithms, it is not everywhere differentiable. However, it should be noted that in our proposed algorithm, similar to the other works in the literature, differentiability is a sufficient condition for the convergence proof. In addition, the probability of meeting the non-differentiable points is almost zero, in practice.}}  \cite{pegasos}: 
\begin{equation}\label{eq: SVM_prob_unconst}
\min_{\boldsymbol{w}} \dfrac{\lambda}{2}  \parallel\boldsymbol{w}  \parallel^2 + \dfrac{1}{m} \sum_{i=1}^{m}  \max \{ 0, 1- y_i \langle \boldsymbol{x}_i , \boldsymbol{w} \rangle  \}  , 
\end{equation} 
where $ \left\lbrace \left( \boldsymbol{x}_i, y_i \right) ,~ \forall i = 1,\ldots, m  \right \rbrace $ is the set of  training samples, and
$ \lambda $ is the regularisation parameter to control overfitting by maintaining a trade-off between increasing the size of the margin (larger $ \lambda $) and ensuring that the data lie on the correct side of the margin (smaller $ \lambda $). 
Once this problem is solved, the obtained  $ {\boldsymbol {w}} $ determines the SVM classifier as $ {\displaystyle {\boldsymbol {x}}\mapsto \operatorname {sgn}( \langle {\boldsymbol {w}} \cdot {\boldsymbol {x}} \rangle} ) $, i.e., each future datum $ \boldsymbol {x} $ will be labelled by the sign of its inner product with the solution $ \boldsymbol {w} $.

Although SVM problem formulation is well understood, solving large-scale SVMs is still challenging. 
For the large and high-dimensional datasets that we encounter in   emerging applications, the optimization problem cast by SVMs 
is huge. Consequently, solving such large-scale SVMs is mathematically complex and computationally expensive. Specifically, as the number of training data becomes very large, computing the exact gradient of the objective function becomes impractical. 
{\color{black}{Therefore, for training large-scale SVMs, off-the-shelf optimization algorithms 
for general problems 
quickly become intractable in their memory 
 requirements.}} 
Moreover, for emerging applications that need to handle a huge amount of data 
 in a short time, the speed of convergence of the applied algorithm is another critical factor, which needs to be carefully considered. 
Furthermore, 
when the data contain a large number of attributes, extensive resources may be required to process the datasets and solve the associated problems. To address this issue especially in resource-limited multi-agent systems, 
it is beneficial to decompose the attributes of the dataset into small groups, each associated with a computing agent, and then, process these smaller datasets by distributed agents in parallel. 

The aforementioned requirements 
give rise to an increasing need for an efficient and scalable method that can solve large-scale SVMs with low complexity and fast convergence. 
In the sequel, we show 
how our proposed algorithm can effectively address the aforementioned issues to solve large-scale SVMs 
with low complexity and faster convergence than the state-of-the-art existing solutions.


First, note that in SVMs with a huge number of training data, 
the optimization problem cast by SVMs (as in \eqref{eq: SVM_prob_unconst}) can be viewed as the sample average approximation (SAA) of the following stochastic optimization problem:
\begin{equation}\label{eq: stoch_svm_prob}
\min_{\boldsymbol{w}} f \left( \boldsymbol{w} \right) = \mathbb{E}_{ \left( \boldsymbol{x}, y \right) }  \left[ \dfrac{\lambda}{2}  \parallel\boldsymbol{w}  \parallel^2 +  \max \{ 0, 1- y \langle \boldsymbol{x} , \boldsymbol{w} \rangle  \}   \right],
\end{equation}
in which the randomness comes from random samples $ \left( \boldsymbol{x} , y \right) $. Considering the batch size $ B=1 $, the stochastic gradient of the above objective function is
\begin{align}\label{eq: stoch_grad_svm}
\hat{\boldsymbol{g}} \left(\boldsymbol{w} \right) = \lambda \boldsymbol{w}  -   y_i \boldsymbol{x}_i  \boldsymbol{1} \left(  y_i \langle \boldsymbol{x}_i , \boldsymbol{w} \rangle \leq 1 \right),
\end{align}
where $ \boldsymbol{1} (.) $ is the indicator function and $ \left( \boldsymbol{x}_i , y_i \right) $ is a randomly drawn training sample. 
Note that since the training samples are chosen i.i.d., the gradient of the loss function with respect to any individual sample can be shown to be an unbiased estimate of a gradient of $ f $ \cite{Shamir2013}. Moreover, the variance of such estimation is finite,  
 i.e., $ \mathrm{E} \left[ \parallel \hat{\boldsymbol{g}} \left(\boldsymbol{w} \right) \parallel^2 \right] \leq G $, as shown in \cite{pegasos}. 
 For computing this stochastic gradient at each iteration, only one sample out of the $  m $ training samples is drawn uniformly at random. Accordingly, the cost of computing the gradient at each iteration is lowered to almost $ 1/m $  of the cost of computing the exact gradient. Therefore, by adopting a stochastic approach and utilising the  stochastic gradient in \eqref{eq: stoch_grad_svm} instead of calculating the exact gradient of the SVM problem formulation, 
 the cost of computing gradient at each iteration will become independent of the number of training data, which makes it highly scalable to large-scale SVMs. 
 
 
 The state-of-the-art algorithms for solving SVMs through a stochastic approach can be classified into \textit{first-order methods}, such as Pegasos \cite{pegasos}, 
 and \textit{second-order methods}, such as the Newton-Armijo method \cite{ssvm}. The existing first-order methods can significantly decrease the computational cost per iteration, 
  but they converge very slowly, which is not desirable specially for emerging applications with huge datasets. On the other hand, second-order methods suffer from high complexity, due to the significantly expensive matrix computations and high storage requirement \cite{convex_big_data,Hessian}. 
For example, the Newton-Amijo method \cite{ssvm} needs to solve a matrix equation, which involves matrix inversion, at each iteration. However, in the case of large-scale SVMs in big data classification, the inversion of a curvature matrix or the storage of an iterative approximation of that inversion will be very expensive, 
which makes this method non-practical for large-scale SVMs. 
In addition, this method is an offline method that needs to have all the training data in a full batch to calculate the exact values of the gradient and Hessian matrix at each iteration. This will cause high complexity and computational cost, and makes this method even not applicable to online applications.

In the next section, we will show that our proposed algorithm can effectively solve large-scale SVMs significantly faster  than the existing solutions, and with low complexity. Using some real-world big datasets, we  compare the proposed accelerated method and the aforementioned methods for solving SVMs.


\section{Simulation Results and Discussion}\label{sec: SVM_sim}


In this section, we empirically evaluate the proposed stochastic optimization algorithm and 
demonstrate 
its efficiency for solving large-scale SVMs. 
We compare the performance of the proposed algorithm to three important state-of-the-art baselines in this area: 
\begin{enumerate}
\item Pegasos algorithm \cite{pegasos}, which is 
known to be one of the best state-of-the-art algorithms for solving large-scale SVMs, 
\item The state-of-the-art parallel stochastic optimization algorithm \cite{Daniel2016parallel}, 
\item ADAM algorithm \cite{ADAM_2014}, one of the most popular and widely-used deep learning 
 algorithms
 , which is an adaptive learning rate optimization algorithm that has been designed specifically for training deep neural networks. This algorithm has been recognized as one of the best optimization algorithms for deep learning, and its popularity is growing 
exponentially, according to \cite{karparthy2017peek}.

\end{enumerate}

We perform our simulations and comparison on two popular real-world large datasets with very different feature counts and sparsity. The COV1 dataset classifies the forest-cover  areas in the Roosevelt National Forest of northern Colorado identified from cartographic variables into two classes \cite{COV1dataset2002}, and the RCV1 dataset  classifies the CCAT and ECAT classes versus GCAT and MCAT classes in the Reuters RCV1 collection \cite{RCV1dataset2004}.\footnote{The datasets are available online at https://www.csie.ntu.edu.tw/~cjlin/libsvmtools/datasets/binary.html.} Details of the datasets' characteristics as well as the values of the SVM regularisation parameter $ \lambda $ used in the experiments are all provided in Table \ref{Datasets for SVM}  
(Note that for the regularisation parameter for each dataset, we  adopt the typical value  used in the previous works \cite{Shamir12, Shamir2013, pegasos} and \cite{linearSVM}). 

Similar to \cite{Shamir2013}, the initial point of all the algorithms is set to $ \boldsymbol{w}_1=\boldsymbol{1} $. 
We  also tune the step-size parameters of the method in \cite{Daniel2016parallel} to achieve its best performance for the used datasets. 
Moreover, for the Pegasos algorithm, we output the last weight vector rather than the average weight vector, as it is found that  it performs better in practice \cite{pegasos}. Finally, for the hyper-parameters of the ADAM algorithm, we use their default values as reported in \cite{ADAM_2014}, which are widely used for almost all the problems and known to perform very well, in practice.

\begin{table}
\caption{The datasets characteristics and the values of the regularisation parameter $ \lambda $ used in the experiments.}\label{Datasets for SVM}
\begin{center}
\begin{tabular}{|  l | *{4}{c}| c  |}
  \hline			
  Dataset & Training Size & Test Size & Features & Sparsity ($ \% $) & $ \lambda $ \\ \hline
  COV1 & 522911 & 58101 & 55 & 22.22 & $ 10^{-6} $ \\
  RCV1 & 677399 & 20242 & 47236 & 0.16 & $ 10^{-4} $  \\
  \hline  
\end{tabular}
\end{center}
\end{table}



\begin{figure*}[]
\begin{center}
\begin{minipage}[]{0.48\linewidth}
\centering
\subfigure[]{
\includegraphics[width=\textwidth]{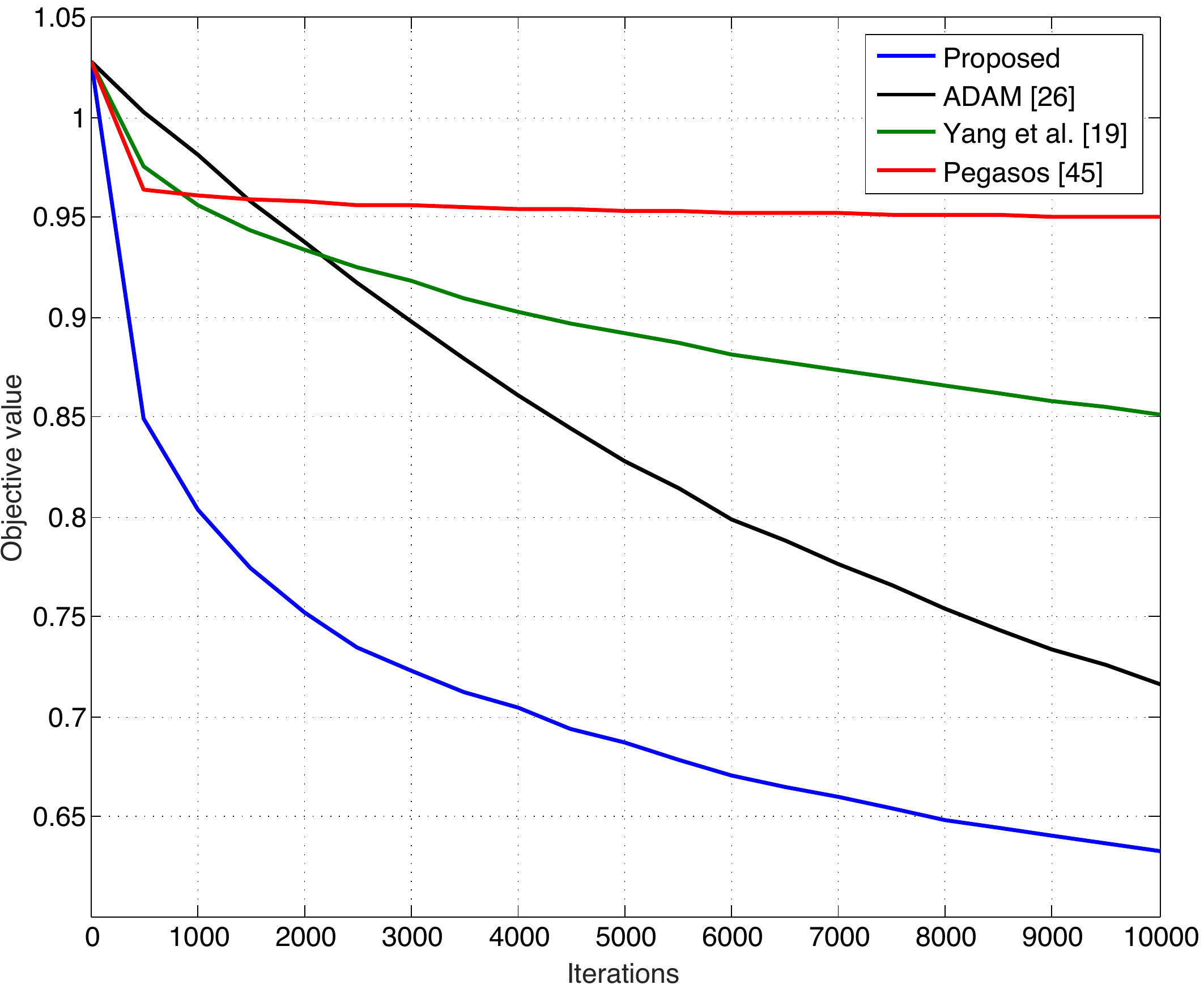}
}
\end{minipage}%
\begin{minipage}[]{0.48\linewidth}
\centering
\subfigure[]{
\includegraphics[width=\textwidth]{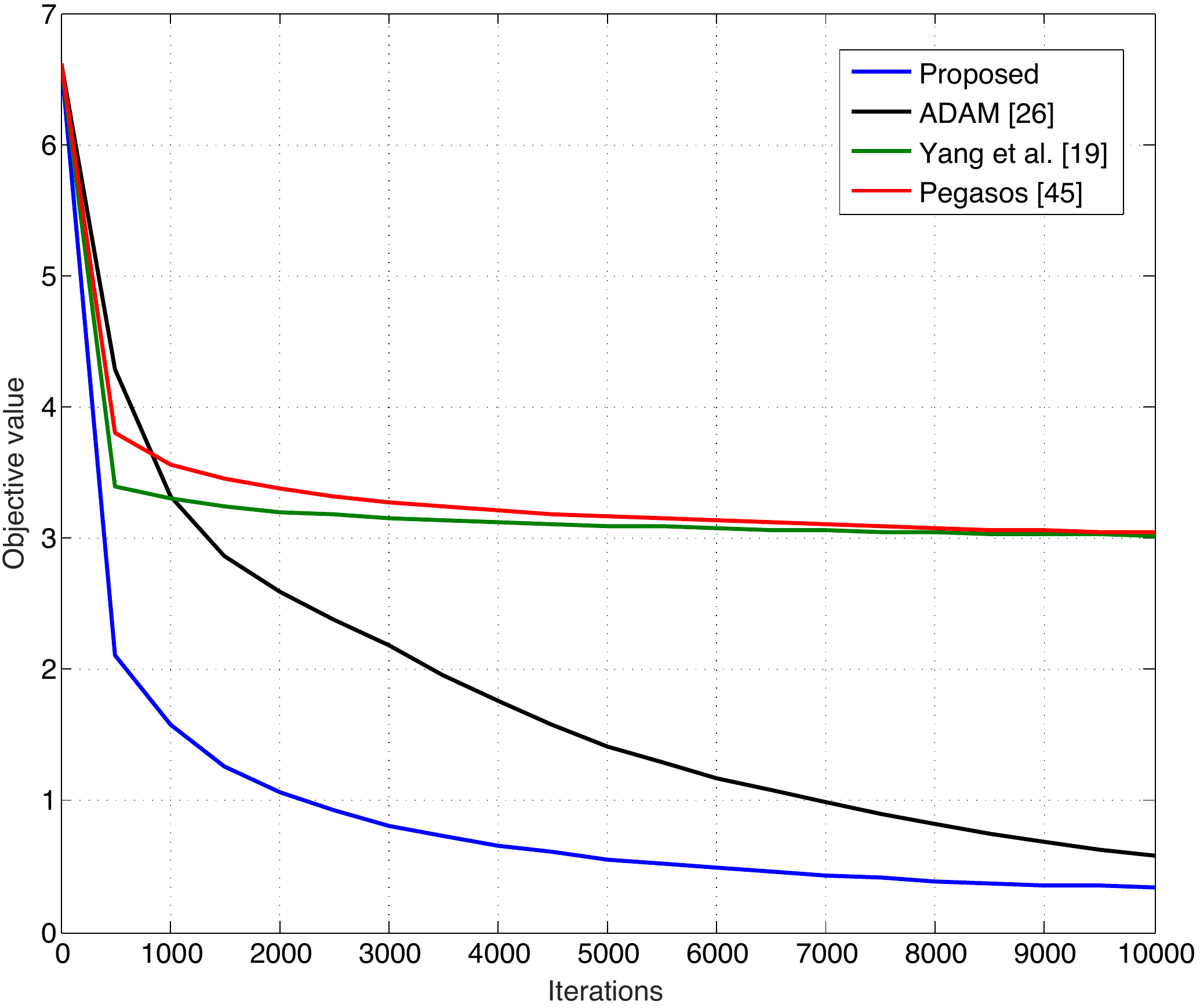}
}
\end{minipage}%
\caption{Comparison of the convergence speed of different methods, on the COV1 (left) and RCV1 (right) datasets.}
\label{SVM_objective_func_Datasets}
\end{center}
\end{figure*}

Fig.s \ref{SVM_objective_func_Datasets}-(a) and \ref{SVM_objective_func_Datasets}-(b) show the convergence results of 
the proposed method and the baselines for each of the datasets. As can be seen in these figures, the proposed algorithm converges much faster than the baselines. Especially at the early iterations, the convergence speeds of all of these methods are rather fast. However, after a number of iterations, the speed of convergence in the baselines drops, while our method  still maintains a good convergence speed. This is because the estimation of the gradient of the objective function  at each iteration in those methods is not as good as in our proposed method. For Pegasos method, this is due to the fact that   the gradient estimation is done based on only one data sample and its associated observed sample function. However in our proposed method, we utilize all the  data samples observed up to that point and 
average over the previous gradients of the observed sample functions to estimate the true gradient of the original objective function. Moroever, the method in \cite{Daniel2016parallel} performs an extra averaging over the iterates, 
which makes the solution change increasingly slowly, as the iterations go on.

\begin{figure*}[]
\begin{center}
\begin{minipage}[]{0.48\linewidth}
\centering
\subfigure[]{
\includegraphics[width=\textwidth]{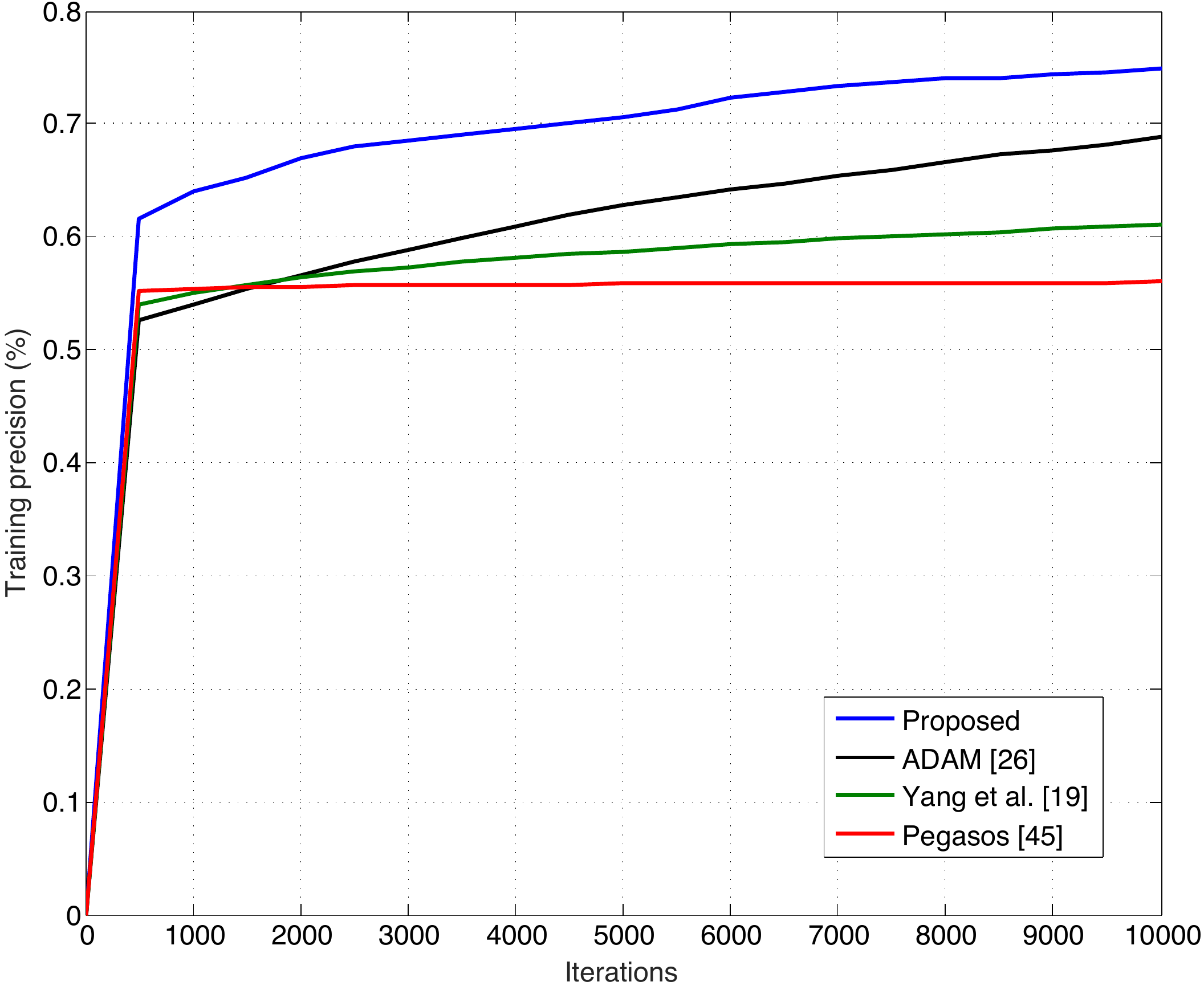}
}
\end{minipage}%
\begin{minipage}[]{0.48\linewidth}
\centering
\subfigure[]{
\includegraphics[width=\textwidth]{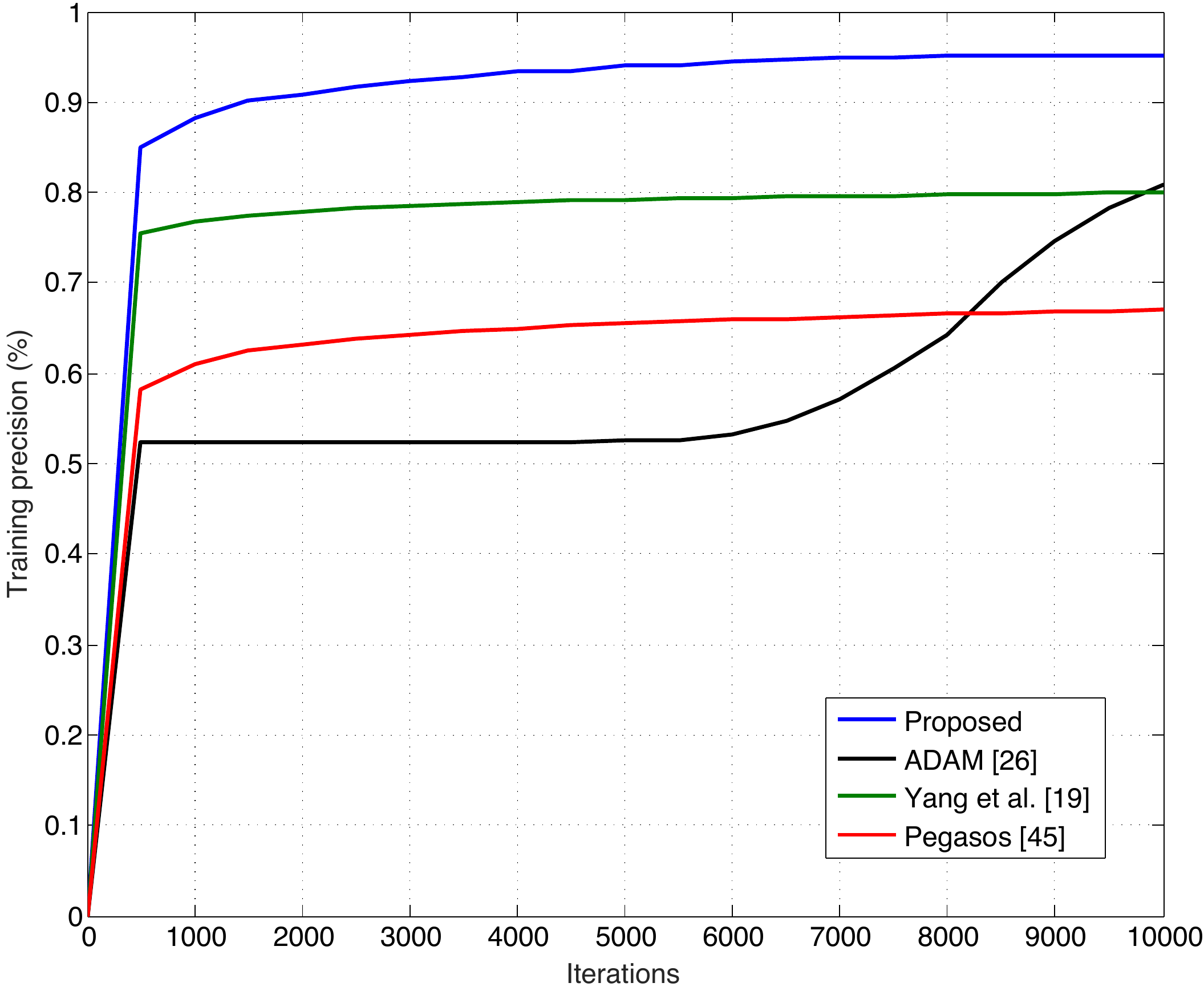}
}
\end{minipage}%
\end{center}
\caption{The percentage of accuracy of the obtained SVM under different methods on the training data of the COV1 (left) and RCV1 (right) datasets.}
\label{SVM_accuracy_Datasets_1}
\end{figure*}

\begin{figure*}[]
\begin{center}
\begin{minipage}[]{0.48\linewidth}
\centering
\subfigure[]{
\includegraphics[width=\textwidth]{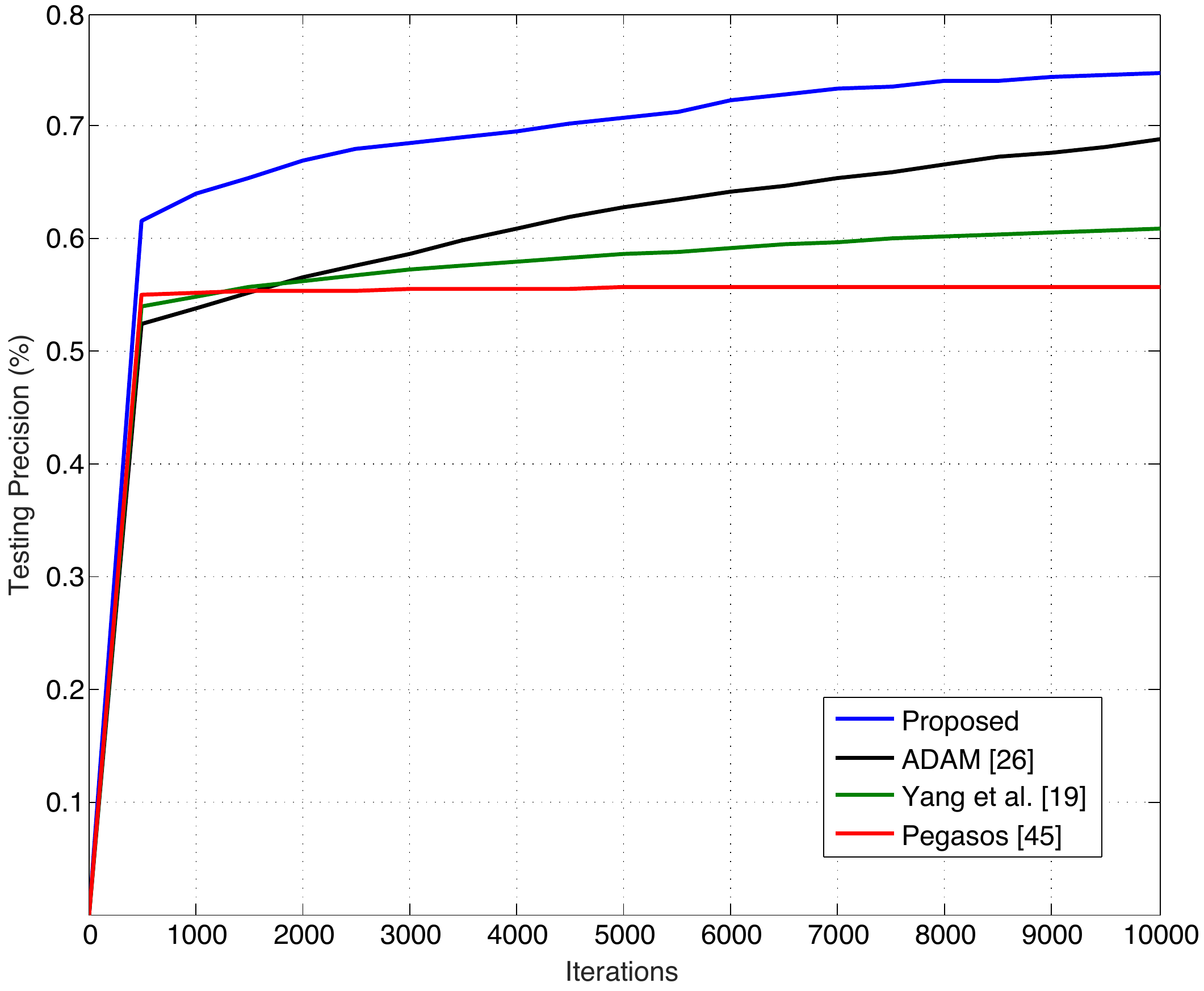}
}
\end{minipage}%
\begin{minipage}[]{0.48\linewidth}
\centering
\subfigure[]{
\includegraphics[width=\textwidth]{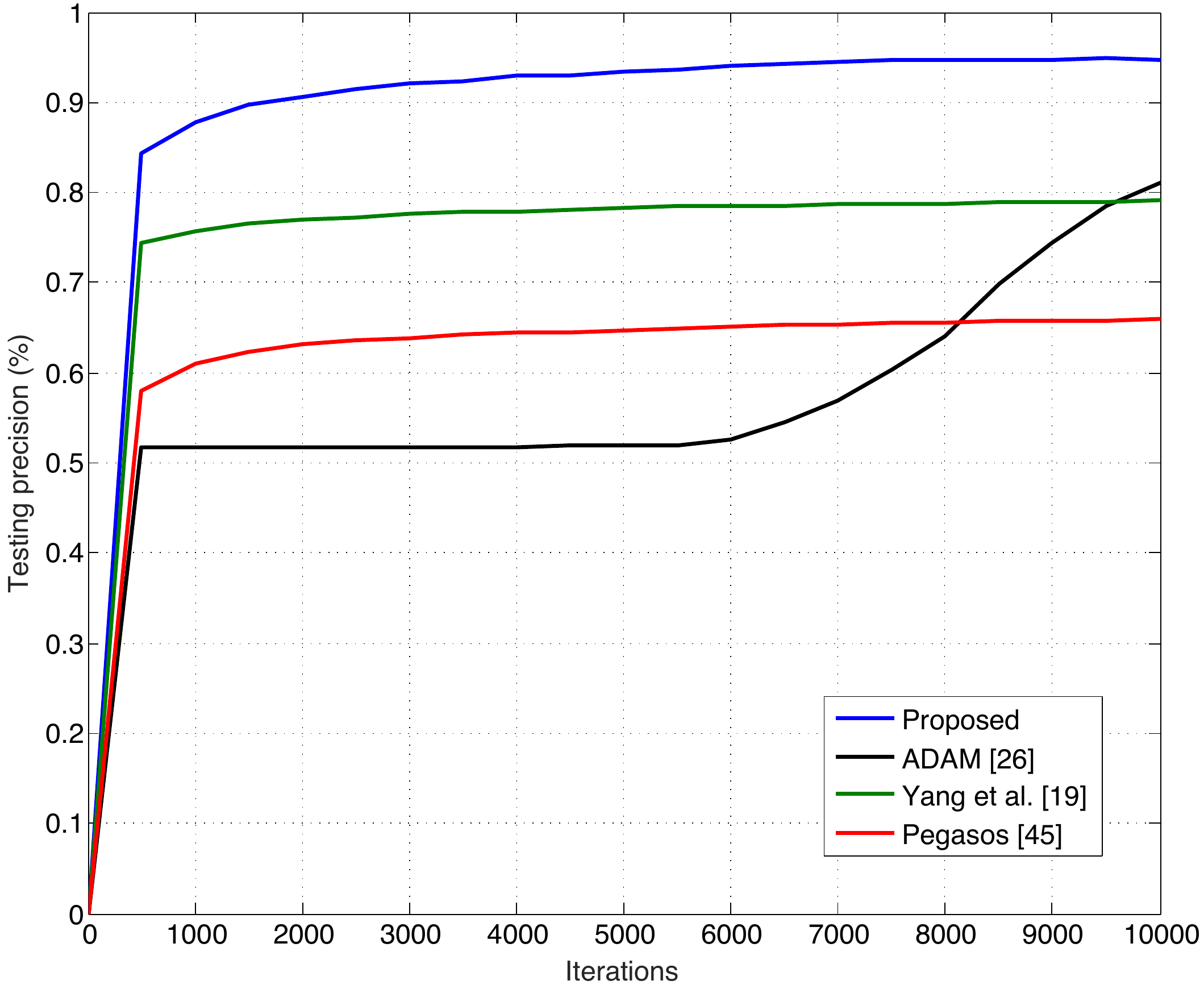}
}
\end{minipage}%
\caption{The percentage of accuracy of the obtained SVM under different methods on the testing data of the COV1 (left) and RCV1 (right) datasets.}
\label{SVM_accuracy_Datasets_2}
\end{center}
\end{figure*}

Figs. \ref{SVM_accuracy_Datasets_1} and \ref{SVM_accuracy_Datasets_2} show the classification precision of the resulted SVM model versus the number of samples visited. The top row shows the classification precision on the training data and the bottom row shows the classification precision on the testing data. 
%
According to the figures, comparing the proposed method to the baselines, under the same number of iterations, the proposed method obtains a more precise SVM with a higher percentage of accuracy for the training and testing data. 

Finally, Table \ref{tab: SVM CPU time} compares the CPU time for running the algorithms for $ 10^4 $ iterations. As can be verified from this table, the CPU time of the proposed method is less than 
 those of the baselines. 
 Therefore, 
the computational complexity per iteration of the proposed algorithm is 
less than the considered baselines, which are known to have low complexity.
\begin{table}
\caption{The CPU time (in seconds) for running $ 10^4 $ iterations of different methods.}\label{tab: SVM CPU time}
\begin{center}
\begin{tabular}{|  l || c | c | c | c |}
  \hline			
  Datasets & Proposed Method & Pegasos \cite{pegasos} & Yang et. al. \cite{Daniel2016parallel} & ADAM \cite{ADAM_2014} \\ \hline \hline
  COV1 & 1.033 & 1.198 & 1.117 & 4.577 \\ \hline
  RCV1 & 73.559 & 74.855 & 74.322  & 84.857 \\
  \hline  
\end{tabular}
\end{center}
\end{table}



\section{Conclusion}\label{sec: SVM_conclusion}

In this paper, we proposed a novel fast parallel stochastic optimization framework that can solve a large class of possibly non-convex constrained stochastic optimization problems. 
Under the proposed method, each user of a multi-agent system updates its control variable in parallel, by solving a   successive convex approximation sub-problem, independently. The sub-problems have low complexity and are easy to obtain. The proposed algorithm can be applied to solve a large class of optimization problems arising in important applications from various fields, such as wireless networks and large-scale machine learning. Moreover, for convex 
problems, we proved the convergence of the proposed algorithm to the optimal solution, and for general non-convex
problems, we proved the convergence of the proposed algorithm to a stationary point.




Moreover, as a representative application of our proposed stochastic optimization framework in the context of machine learning, we elaborated on large-scale SVMs 
and demonstrated how the proposed algorithm can efficiently solve this problem
  , especially for modern applications with huge datasets. 
We compared the performance of our proposed algorithm to
 the state-of-the-art baselines. Numerical results on popular real-world datasets show that the proposed method can significantly outperform the state-of-the-art 
 methods in terms of the convergence speed while having the same or lower complexity and storage requirement.

\bibliography{Bibliography}
\bibliographystyle{IEEEtran}


\appendices

\section{Convergence of the Proposed Algorithm}\label{sec: Proofs}

In order to prove 
Theorems \ref{th: conv_nonconvex}-\ref{th: conv2_nonconvex}, we first present 
five useful lemmas as follows.

\begin{lemma}\label{lem: h_k sum update}
Under Algorithm \ref{alg: main}, for any $ l=1,\cdots,I $, we have
\begin{equation}\label{eq: h_k sum update}
\forall k \geq 1: \quad \boldsymbol{h}^k_l = \sum_{i=1}^{k} \omega_k^i \nabla_l f^i \left( \boldsymbol{x}_l^{i-1} \right) ,
\end{equation}
\end{lemma}
where the weights $ \omega_k^i , \forall i=1,\cdots,k $ are defined as
\begin{equation}\label{eq: omega_k^i}
\omega_k^i \triangleq \omega_i \prod_{j=i+1}^k \left( 1-\omega_j \right).
\end{equation}

\proof ~
The proof follows from a simple induction by applying the update equation of $ \boldsymbol{h}_l^k , ~\forall k $ recursively, as follows. For any fixed agent index $ l $, starting from $ k=1 $ and using the update equation in \eqref{eq: h^k update} along with the initial values of $ \boldsymbol{h}_l^0 = \boldsymbol{0} $ and $ \omega_1=1 $, we have $ \boldsymbol{h}_l^1 = \nabla_l f^1 \left( \boldsymbol{x}_l^{0} \right) $, which satisfies the form introduced in \eqref{eq: h_k sum update}. 
Now assume that for some $ k \geq 1 $, $ \boldsymbol{h}^k_l = \sum_{i=1}^{k} \omega_k^i \nabla_l f^i \left( \boldsymbol{x}_l^{i-1} \right) $. It suffices to show that $ \boldsymbol{h}^{k+1}_l = \sum_{i=1}^{k+1} \omega_{k+1}^i \nabla_l f^i \left( \boldsymbol{x}_l^{i-1} \right) $, as well. 
%
%

Using the update equation \eqref{eq: h^k update} for the $ \mathrm{(k+1)}^{st} $ iteration, we have
\begin{equation}
\boldsymbol{h}^{k+1}_l = \left( 1- \omega_{k+1} \right) \boldsymbol{h}^{k}_l + \omega_{k+1} \nabla_l f^{k+1} \left( \boldsymbol{x}_l^{k} \right).
\end{equation}
Substituting $  \boldsymbol{h}^{k}_l $, in the above equation results in
\begin{align}
\boldsymbol{h}^{k+1}_l &= \left( 1- \omega_{k+1} \right) \sum_{i=1}^{k} \omega_k^i \nabla_l f^i \left( \boldsymbol{x}_l^{i-1} \right) + \omega_{k+1} \nabla_l f^{k+1} \left( \boldsymbol{x}_l^{k} \right) \notag \\
&=\sum_{i=1}^{k} \omega_{k+1}^i \nabla_l f^i \left( \boldsymbol{x}_l^{i-1} \right) + \omega_{k+1} \nabla_l f^{k+1} \left( \boldsymbol{x}_l^{k} \right) \notag \\
&=\sum_{i=1}^{k+1} \omega_{k+1}^i \nabla_l f^i \left( \boldsymbol{x}_l^{i-1} \right),
\end{align}
where the second equality follows from the fact that according to the definition of $ \omega_k^i $ in \eqref{eq: omega_k^i}, $  \left(1-\omega_{k+1} \right)  \omega_k^i = \omega_{k+1}^i , ~ \forall k \geq 1, 1\leq \forall  i \leq k $.  $ \hfill \blacksquare $

\begin{lemma}\label{lem: h_k bounded}
Under Algorithm \ref{alg: main} and for any $ k \geq 1 $,  $ \parallel  \boldsymbol{h}^k_l \parallel  , ~ \forall l $  is bounded. 
\end{lemma}

\proof ~
 The proof is a direct result of Lemma \ref{lem: h_k sum update}. Since according to Assumption \ref{assump: stoch opt formulation}-b,  $ \nabla_l f^i \left( \boldsymbol{x}\right), \forall i  $ are Lipschitz continuous over the domain $ \mathcal{X} $, it follows that there exists some $ 0 \leq G \leq \infty $ such that $ \nabla_l f^i \left( \boldsymbol{x}\right), \forall i, \forall \boldsymbol{x} \in  \mathcal{X} $. Therefore, using \eqref{eq: h_k sum update}, we have 
\begin{equation}\label{eq: h_k bounded_v2}
\parallel \boldsymbol{h}^k_l \parallel = \parallel \sum_{i=1}^{k} \omega_k^i \nabla_l f^i \left( \boldsymbol{x}_l^{i-1} \right) \parallel \leq \sum_{i=1}^{k} \omega_k^i G.
\end{equation}
Moreover, from \eqref{eq: omega_k^i} and the assumption that $ \omega_1 = 1 $, it is easy to verify that 
\begin{equation}\label{eq: sum w_k^i =1}
\sum_{i=1}^{k} \omega_k^i  = 1, \quad \forall k \geq 1.
\end{equation}
Combining \eqref{eq: h_k bounded_v2} and \eqref{eq: sum w_k^i =1}  results in 
\begin{equation}
\parallel \boldsymbol{h}^k_l \parallel  \leq G,
\end{equation}
which completes the proof. $ \hfill \blacksquare $

\begin{lemma}\label{lem: x^k - x^k-1}
Under Algorithm \ref{alg: main}, for any $ l=1,\cdots,I $, we will have $  \parallel \boldsymbol{x}_l^k - \boldsymbol{x}_l^{k-1}  \parallel  = O \left( \alpha_k \right) $, and hence, 
\begin{equation}
\lim_{k \rightarrow \infty} \parallel \boldsymbol{x}^k_l - \boldsymbol{x}^{k-1}_l  \parallel = 0, \quad \forall l=1,\cdots,I.   
\end{equation}
\end{lemma}

\proof  ~
For any $ l $, since 
 $ \boldsymbol{x}^k_l $ is the minimiser of $ \hat{f}_l^k(\boldsymbol{x}_l) $ and $ \boldsymbol{x}^{k-1}_l $ is a feasible point,  we have 
 \begin{align}
 & \hat{f}_l^k(\boldsymbol{x}_l^k) \leq \hat{f}_l^k(\boldsymbol{x}_l^{k-1})  \notag \\
\Rightarrow ~ &\dfrac{1}{2 \alpha_k} \parallel \boldsymbol{x}_l^k - \boldsymbol{x}_l^{k-1}  \parallel^2 + \langle \boldsymbol{h}^k_l , \boldsymbol{x}_l^k - \boldsymbol{x}_l^{k-1}  \rangle  \leq 0, \notag \\
 \Rightarrow ~ & \dfrac{1}{2 \alpha_k} \parallel \boldsymbol{x}_l^k - \boldsymbol{x}_l^{k-1}  \parallel^2 - \parallel \boldsymbol{h}^k_l \parallel . \parallel \boldsymbol{x}_l^k - \boldsymbol{x}_l^{k-1} \parallel  \leq 0, \notag \\ 
 \Rightarrow ~ &   \parallel \boldsymbol{x}_l^k - \boldsymbol{x}_l^{k-1}  \parallel \leq 2 \alpha_k \parallel \boldsymbol{h}^k_l \parallel .
 \end{align}
 Therefore, since according to Lemma \ref{lem: h_k bounded}, $ \parallel \boldsymbol{h}^k_l \parallel  $ is bounded, we have $  \parallel \boldsymbol{x}_l^k - \boldsymbol{x}_l^{k-1}  \parallel  = O \left( \alpha_k \right) $. Consequently, as $ \alpha_k $ is decreasing (due to Assumption \ref{assump: step-sizes}-a), $ \lim_{k \rightarrow \infty} \parallel \boldsymbol{x}^k_l - \boldsymbol{x}^{k-1}_l  \parallel = 0 $.  $ \hfill\blacksquare $


\begin{lemma}\label{lem: h_k  --> nabla f}
Under Algorithm \ref{alg: main}, 
the vector $ \boldsymbol{h}^k $ converges to the true gradient of the objective function
, with probability one, i.e., 
\begin{equation}
\lim_{k \rightarrow \infty} \parallel \boldsymbol{h}^k - \nabla F \left( \boldsymbol{x}^k \right) \parallel = 0, \quad \mathrm{w.p.1}.
\end{equation}
\end{lemma}

\proof ~
We simply refer to \cite[Lemma 1]{ruszczynski1980feasible}, and show that all of the required conditions of that lemma are satisfied in our case. Specifically, Conditions (a) and (b) of \cite[Lemma 1]{ruszczynski1980feasible} are satisfied due to Assumption \ref{assump: stoch opt formulation}-a and Assumption \ref{assump: noise}, respectively. Moreover, Conditions (c) and (d) of \cite[Lemma 1]{ruszczynski1980feasible} are satisfied by Assumptions \ref{assump: step-sizes} and \ref{assump: noise}-a. Finally, it remains to prove that Condition (e) of \cite[Lemma 1]{ruszczynski1980feasible} is satisfied under our problem as well, which will be shown in the following.

Note that since the objective function has a Lipschitz continuous gradient (according to Assumption \ref{assump: stoch opt formulation}), we have
\begin{equation} \label{eq: Condition e}
\frac{\parallel f \left( \boldsymbol{x}_l^k \right) - f \left( \boldsymbol{x}_l^{k-1} \right)  \parallel }{\omega_k} \leq  L_{\nabla F} \frac{\parallel \boldsymbol{x}_l^k - \boldsymbol{x}_l^{k-1} \parallel }{\omega_k} .
\end{equation}
Moreover, due to Lemma \ref{lem: x^k - x^k-1}, we have $  \parallel \boldsymbol{x}_l^k - \boldsymbol{x}_l^{k-1}  \parallel  = O \left( \alpha_k \right) $. The above two facts along with Assumption \ref{assump: step-sizes} on the sequences $ \left\lbrace \omega_k \right\rbrace $ and $ \left\lbrace \alpha_k \right\rbrace $ concludes that the right-hand side of inequality \eqref{eq: Condition e} goes to zero as $ k $ goes to infinity, and hence,
\begin{equation}
\lim_{k \rightarrow \infty}  \parallel f \left( \boldsymbol{x}_l^k \right) - f \left( \boldsymbol{x}_l^{k-1} \right)  \parallel  / \omega_k = 0.
\end{equation}
Therefore, it follows from 
\cite[Lemma 1]{ruszczynski1980feasible} that $ \boldsymbol{h}^k - \nabla f \left( \boldsymbol{x}^k \right) \rightarrow 0 $, w.p.1. $ \hfill \blacksquare $

Now, with the above lemmas, we proceed to prove Theorem \ref{th: conv_convex}.
First of all, since the sequence 
 $ \left\lbrace \boldsymbol{x}^k \right\rbrace $ lies in a compact set, it is sufficient to show that every limit point of the iterates is a stationary point of Problem \ref{eq: main stochastic optimisation}.
To show this, let $ \bar{\boldsymbol{x}} $ be the limit point of a convergent subsequent  $ \left\lbrace \boldsymbol{x}^{k_j} \right\rbrace_{j=1}^\infty $. Note that since $ \mathcal{X} $ is a closed set, this limit point belongs to this set, and hence, it is a feasible point to the optimization problem \ref{eq: main stochastic optimisation}.

\begin{lemma}\label{lem: lim  |x_k -x_k-1| /alfa_k=0} 
There exists a subsequence $ \left\lbrace k_j \right\rbrace_{j=1}^{\infty} $ such that under the proposed Algorithm \ref{alg: main}, 
\begin{align}
\lim_{j \rightarrow \infty} \dfrac{ \parallel \boldsymbol{x}^{k_j} - \boldsymbol{x}^{{k_j}-1} \parallel}{\alpha_{k_j}} = 0. 
\end{align}
\end{lemma}

\proof ~
 We prove this lemma by contradiction. Assume that the statement of the lemma is not true (i.e., there is no such subsequence). This means that for any $ \epsilon > 0 $, there exists some $ k_\epsilon >0 $ such that 
\begin{align}\label{eq: |x_k -x_k-1| /alfa_k > eps}
\forall k \geq k_\epsilon: \quad   \dfrac{ \parallel \boldsymbol{x}^{k} - \boldsymbol{x}^{{k}-1} \parallel}{\alpha_{k}}  \geq \epsilon.
\end{align}

Furthermore, note that from the first-order optimality conditions of \eqref{eq: x^k update}, we have
\begin{align}
 \langle \boldsymbol{x}_l - \boldsymbol{x}^k_l , \dfrac{1}{\alpha_k} \left( \boldsymbol{x}_l - \boldsymbol{x}_l^k \right) + \boldsymbol{h}^k_l \rangle \geq 0, \quad \forall k \geq 1, \forall l, \forall \boldsymbol{x}_l \in \mathcal{X}_l,
\end{align}
which, by substituting $ \boldsymbol{x}_l = \boldsymbol{x}^{k-1}_l $, leads to
\begin{align}\label{eq: < x_k - x_k-1 , h_k > leq ... }
 \langle \boldsymbol{x}^k_l - \boldsymbol{x}^{k-1}_l , \boldsymbol{h}^k_l \rangle \leq  -\dfrac{\parallel \boldsymbol{x}^k_l - \boldsymbol{x}_l^{k-1} \parallel^2 }{\alpha_k}   ,   \quad \forall k \geq 1, \forall l.
\end{align}
Moreover, it follows from Assumption \ref{assump: stoch opt formulation}-c that
\begin{align}\label{eq: f_k - f_k-1<..._v0}
f \left( \boldsymbol{x}^k \right) - f \left( \boldsymbol{x}^{k-1} \right) & \leq \langle \boldsymbol{x}^k - \boldsymbol{x}^{k-1} , \nabla f \left( \boldsymbol{x}^{k-1} \right) \rangle + \frac{L_{\nabla f}}{2} \parallel \boldsymbol{x}^k - \boldsymbol{x}^{k-1} \parallel^2 \notag \\
& \leq \langle \boldsymbol{x}^k - \boldsymbol{x}^{k-1} , \nabla f \left( \boldsymbol{x}^{k-1} \right)  - \boldsymbol{h}^k + \boldsymbol{h}^k \rangle + \frac{L_{\nabla f}}{2} \parallel \boldsymbol{x}^k - \boldsymbol{x}^{k-1} \parallel^2 \notag \\
& \leq \parallel \boldsymbol{x}^k - \boldsymbol{x}^{k-1} \parallel \parallel \nabla f \left( \boldsymbol{x}^{k-1} \right)  - \boldsymbol{h}^k \parallel +\langle \boldsymbol{x}^k - \boldsymbol{x}^{k-1} ,  \boldsymbol{h}^k \rangle \notag \\
& \quad + \frac{L_{\nabla f}}{2} \parallel \boldsymbol{x}^k - \boldsymbol{x}^{k-1} \parallel^2. 
\end{align}
Plugging the inequalities \eqref{eq: |x_k -x_k-1| /alfa_k > eps} and \eqref{eq: < x_k - x_k-1 , h_k > leq ... }  into \eqref{eq: f_k - f_k-1<..._v0} it is obtained that for any $ k \geq k_\epsilon $,
\begin{align}\label{eq: f^k -f^k-1 < ...}
f \left( \boldsymbol{x}^k \right) - f \left( \boldsymbol{x}^{k-1} \right) & \leq - \left( \dfrac{1}{\alpha_k} ( 1- \dfrac{\parallel \nabla f \left( \boldsymbol{x}^{k-1} \right)  - \boldsymbol{h}^k \parallel}{\epsilon} ) - \frac{L_{\nabla f}}{2}   \right)  \parallel \boldsymbol{x}^k - \boldsymbol{x}^{k-1} \parallel^2.
\end{align}

Considering the facts that $ \lim_{k \rightarrow \infty} \parallel \boldsymbol{x}^k - \boldsymbol{x}^{k-1}  \parallel = 0 $ and $ \lim_{k \rightarrow \infty} \parallel \boldsymbol{h}^k - \nabla f \left( \boldsymbol{x}^k \right) \parallel = 0, ~ \mathrm{w.p.1} $ (according to Lemmas \ref{lem: x^k - x^k-1} and \ref{lem: h_k --> nabla f}, respectively) and the assumption that $ \left\lbrace \alpha_k \right\rbrace $ is a decreasing sequence (due to Assumption \ref{assump: step-sizes}), it follows that there exists a sufficiently large $ k^\prime_\epsilon \geq k_\epsilon $ such that for any $ k \geq k^\prime_\epsilon $, we have 
\begin{equation}\label{eq: ... > lamda/alfa}
\dfrac{1}{\alpha_k} ( 1- \dfrac{\parallel \nabla f \left( \boldsymbol{x}^{k-1} \right)  - \boldsymbol{h}^k \parallel}{\epsilon} ) - \frac{L_{\nabla f}}{2}  \geq  \dfrac{\lambda_\epsilon}{\alpha_k}
\end{equation}
for some $ \lambda_\epsilon >0 $, with probability one. Therefore, it follows from \eqref{eq: f^k -f^k-1 < ...} that for any $ k \geq k^\prime_\epsilon $, 
\begin{align}\label{eq: f^k -f^k-1 < ..._v2}
f \left( \boldsymbol{x}^k \right) - f \left( \boldsymbol{x}^{k-1} \right) & \leq - \dfrac{\lambda_\epsilon}{\alpha_k} \parallel \boldsymbol{x}^k - \boldsymbol{x}^{k-1} \parallel^2, \quad \mathrm{w.p.1},
\end{align}
and hence, invoking \eqref{eq: |x_k -x_k-1| /alfa_k > eps}, we have
\begin{align}\label{eq: f^k -f^k-1 < ..._v3}
\forall \epsilon >0, \forall k > k^\prime_\epsilon: \quad f \left( \boldsymbol{x}^k \right) - f \left( \boldsymbol{x}^{k^\prime_\epsilon} \right) \leq - \epsilon^2 \lambda_\epsilon \sum_{i= k^\prime_\epsilon}^{k} \alpha_i , \quad \mathrm{w.p.1}.
\end{align}
Letting $ k \rightarrow \infty $, the right-hand side of \eqref{eq: f^k -f^k-1 < ..._v3} will go to $ - \infty $, as $ \sum_{i= k^\prime_\epsilon}^{\infty} \alpha_i = \infty $ according to Assumption \ref{assump: step-sizes}. This contradicts the boundedness of the objective function $ f \left( \boldsymbol{x} \right) $ over the domain $ \mathcal{X} $. Therefore, 
the contradiction assumption 
 in \eqref{eq: |x_k -x_k-1| /alfa_k > eps} must be wrong and this completes the proof. $ \hfill \blacksquare $

Now, having the above useful results, in the rest of this section, we will prove the main convergence results stated in Theorems \ref{th: conv_convex}--\ref{th: conv2_nonconvex}. In the following, we first prove Theorem \ref{th: conv_nonconvex}, which is for the general (convex or non-convex) case, and then  use it in the proofs of Theorems \ref{th: conv_convex} and \ref{th: conv2_nonconvex} as well.

\subsection{Proof of Theorem \ref{th: conv_nonconvex}}

First, note that according to Lemma \ref{lem: lim  |x_k -x_k-1| /alfa_k=0}, there exists a subsequence $ \left\lbrace k_j \right\rbrace_{j=0}^{\infty} $ such that 

$ \hspace{-10 pt} \displaystyle\lim_{j \rightarrow \infty} \dfrac{ \parallel \boldsymbol{x}^{k_j} - \boldsymbol{x}^{{k_j}-1} \parallel}{\alpha_{k_j}} = 0 $. Let $ \bar{\boldsymbol{x}}=\left( \bar{\boldsymbol{x}}_l \right)_{l=1}^I $ be the limit point of the convergent subsequence $ \left\lbrace \boldsymbol{x}^{k_j} \right\rbrace $. Since $ \mathcal{X} $ is a closed and convex set, we have $ \bar{\boldsymbol{x}} \in \mathcal{X} $. In the following, we will show that this limit point is a stationary point for the stochastic optimization problem in \eqref{eq: main stochastic optimisation}.

Invoking the first-order optimality condition of \ref{eq: x^k update}, we have $ \forall l=1,\ldots,I, \forall \boldsymbol{x}_l \in \mathcal{X}_l $, 
\begin{align}\label{eq: first-order opt cond_v3}
& \lim_{j \rightarrow \infty}  \langle \boldsymbol{x}_l - \boldsymbol{x}_l^{k_j} , \nabla_l \hat{f}_l^{k_j} \left( \boldsymbol{x}_l^{k_j} \right) \rangle \notag \\
&~~ = \lim_{j \rightarrow \infty}  \langle \boldsymbol{x}_l - \boldsymbol{x}_l^{k_j} , \dfrac{1}{\alpha_{k_j}}  \left( \boldsymbol{x}_l^{k_j} - \boldsymbol{x}_l^{k_j-1} \right) + \boldsymbol{h}_l^{k_j} \rangle \notag \\
&~~ = \lim_{j \rightarrow \infty}  \langle \boldsymbol{x}_l - \boldsymbol{x}_l^{k_j} , \dfrac{1}{\alpha_{k_j}}  \left( \boldsymbol{x}_l^{k_j} - \boldsymbol{x}_l^{k_j-1} \right) + \left( \boldsymbol{h}_l^{k_j} - \nabla_l f \left( \boldsymbol{x}_l^{k_j} \right) \right) + \nabla_l f \left( \boldsymbol{x}_l^{k_j} \right)  \rangle \notag \\
&~~ = \langle \boldsymbol{x}_l -  \bar{\boldsymbol{x}}_l , \nabla_l f \left( \bar{\boldsymbol{x}}_l \right)  \rangle 
\geq  0,
\end{align} 
where the last equality follows from the facts that $ \displaystyle\lim_{j \rightarrow \infty} \dfrac{ \parallel \boldsymbol{x}^{k_j} - \boldsymbol{x}^{{k_j}-1} \parallel}{\alpha_{k_j}} = 0 $ (as shown by Lemma \ref{lem: lim  |x_k -x_k-1| /alfa_k=0}) and $ \lim_{k \rightarrow \infty} \parallel \boldsymbol{h}^k - \nabla f \left( \boldsymbol{x}^k \right) \parallel = 0 $ (as shown by Lemma \ref{lem: h_k  --> nabla f}). Summing \eqref{eq: first-order opt cond_v3} over all $ l=1,\ldots,I $, we obtain 
\begin{equation}
\langle \boldsymbol{x} -  \bar{\boldsymbol{x}} , \nabla f \left( \bar{\boldsymbol{x}} \right)  \rangle \geq 0, \quad \forall \boldsymbol{x} \in \mathcal{X},
\end{equation}
which is the desired first-order optimality condition for the considered optimization problem in \eqref{eq: main stochastic optimisation}. Therefore, the limit point $ \bar{\boldsymbol{x}} $ is a stationary point of \eqref{eq: main stochastic optimisation}. This completes the proof of Theorem \ref{th: conv_nonconvex}.  $ \hfill \square $

 Note that in the case of a convex objective function $ f $, this stationary point is a global minimum, and hence we will have $ f \left( \bar{\boldsymbol{x}} \right)  = f^\ast $, where $ f^\ast $ is the optimal value of the considered optimization problem in \eqref{eq: main stochastic optimisation}.\footnote{Note that this is the first place that makes use of the convexity of $ f $, and all the previous results are applicable to the non-convex case, as well.}

\subsection{Proof of Theorem \ref{th: conv_convex} (Convex case)}

First, for any fixed $ \epsilon > 0 $,  we define the following two sets
\begin{equation}\label{eq: def of X_eps}
\mathcal{X}_\epsilon \triangleq \left\lbrace  \boldsymbol{x}  \in \mathcal{X} \left| F \left( \boldsymbol{x} \right) - F \left( \boldsymbol{x}^\ast \right)  < \epsilon \right. \right\rbrace ,
\end{equation}
\begin{equation}\label{eq: def of bar_X_eps}
\bar{\mathcal{X}_\epsilon}  \triangleq \mathcal{X} - \mathcal{X}_\epsilon,
\end{equation}
and the set of indices
\begin{equation}\label{eq: def of bar_K_eps}
\bar{\mathcal{K}_\epsilon}  \triangleq \left\lbrace  k \left| \boldsymbol{x}^k  \in \bar{\mathcal{X}_\epsilon} \right. \right\rbrace.
\end{equation}

Second, we prove the following lemma, which will then be used in the proof of Theorem \ref{th: conv_convex}.

\begin{lemma}{\textbf{(Convex case)}}\label{lem: convex, k in bar_K_eps}
In the case of the convex objective function in 
\eqref{eq: main stochastic optimisation}, there exists some $ \epsilon^\prime > 0 $ such that under Algorithm \ref{alg: main}, 
we have
\begin{equation}\label{eq: >eps_prime}
\dfrac{\parallel \boldsymbol{x}^k - \boldsymbol{x}^{k-1} \parallel}{\alpha_k} > \epsilon^\prime, \quad  \forall k \in \bar{\mathcal{K}_\epsilon}.
\end{equation}
\end{lemma}

\proof ~
First note that if $ | \bar{\mathcal{K}_\epsilon} | < \infty $, 
then the statement of the lemma is obvious. 
 Therefore, it suffices to prove the lemma for the case where $ | \bar{\mathcal{K}_\epsilon} | = \infty $. We prove this case by contradiction, as follows. 
Suppose that there exists no $ \epsilon^\prime >0 $ that satisfies \eqref{eq: >eps_prime}. Therefore, there should exist some subsequence 
$ \left\lbrace k^\prime_j \right\rbrace \subset \bar{\mathcal{K}_\epsilon} $ such that  
\begin{equation}
\lim_{j \rightarrow \infty}  \dfrac{\parallel \boldsymbol{x}^{k^\prime_j} - \boldsymbol{x}^{k^\prime_j-1} \parallel}{\alpha_{k^\prime_j} } =0.
\end{equation}
This means that the sequence $ \left\lbrace \boldsymbol{x}_{k^\prime_j} \right\rbrace $ is convergent to a limit point. Let $ \bar{\boldsymbol{x}}^\prime=\left( \bar{\boldsymbol{x}}^\prime_l \right)_{l=1}^I $ be the limit point of this convergent subsequence. 
Note that since $ \mathcal{X} $ is a closed and convex set, this limit point belongs to $ \mathcal{X} $, as well. 
 Using a similar analysis to that in \eqref{eq: first-order opt cond_v3}, for the convergent sequence $ \left\lbrace \boldsymbol{x}_{k^\prime_j} \right\rbrace $, 
it is easy to verify that 
\begin{equation}
\langle \boldsymbol{x} -  \bar{\boldsymbol{x}}^\prime , \nabla f \left( \bar{\boldsymbol{x}}^\prime \right)  \rangle \geq 0, \quad \forall \boldsymbol{x} \in \mathcal{X},
\end{equation} 
and hence, the limit point $ \bar{\boldsymbol{x}}^\prime $ is a stationary point to the optimization problem \eqref{eq: main stochastic optimisation}. 
Consequently, since the problem is assumed to be convex, this stationary point is the global minimum, and hence, we have $ F \left( \bar{\boldsymbol{x}}^\prime \right)  = F^\ast $, or equivalently
\begin{equation} 
\displaystyle \lim_{j \rightarrow \infty}  
F \left( \boldsymbol{x}^{k^\prime_j} \right)  
= F^\ast.
\end{equation}
Accordingly, for any fixed $ \gamma > 0 $, there exists some sufficiently large $ k^\prime_\gamma $ such that
\begin{equation}
\forall k^\prime_j \geq k^\prime_\gamma  : \quad F \left( \boldsymbol{x}^{k^\prime_j} \right) - F^\ast < \gamma.
\end{equation}
Substituting $ \gamma = \epsilon $, it follows that 
for sufficiently large $ k^\prime_j \in \bar{\mathcal{K}_\epsilon}  $, we have $ F \left( \boldsymbol{x}^{k^\prime_j} \right) - F^\ast < \epsilon $. This means that $ \boldsymbol{x}^{k^\prime_j} \in \mathcal{X}_\epsilon $, which is obviously a contradiction of the fact that $ k^\prime_j \in \bar{\mathcal{K}_\epsilon} $. 
Therefore, the contradiction assumption must be wrong, and this completes the proof of the lemma. 
 $ \hfill \blacksquare $


Since the subsequence $ \left\lbrace \boldsymbol{x}^{k_j} \right\rbrace $ converges to a stationary point of \eqref{eq: main stochastic optimisation}, 
and the function $ F \left( \boldsymbol{x} \right) $ is continuous (due to Assumption \ref{assump: stoch opt formulation}), 
 it follows that there exists a sufficiently large $ \hat{k}_\epsilon  \in \left\lbrace k_j \right\rbrace_{j=1}^{\infty} $ such that, 
  for any $ k \geq \hat{k}_\epsilon $,  
  $ F \left( \boldsymbol{x}^{\hat{k}_\epsilon} \right)  - F^\ast 
\leq \epsilon $. Consequently, due to the definition in \eqref{eq: def of X_eps}, we have 
\begin{equation}\label{eq: x_hat_k  in  X_eps}
\boldsymbol{x}^{\hat{k}_\epsilon}  \in  \mathcal{X}_\epsilon.
\end{equation}
In the following, we will show by contradiction that for any $ k > \hat{k}_\epsilon  $, $ \boldsymbol{x}^k \in \mathcal{X}_\epsilon $, as well. 

Suppose $ \boldsymbol{x}^{\hat{k}_\epsilon+1}  \notin \mathcal{X}_\epsilon $, and hence, $ \boldsymbol{x}^{\hat{k}_\epsilon+1}  \in \bar{\mathcal{X}_\epsilon} $. Therefore, according to Lemma \ref{lem: convex, k in bar_K_eps}, there exists some fixed  $ \epsilon^\prime > 0 $ such that
\begin{equation}
\dfrac{ \parallel \boldsymbol{x}^{\hat{k}_\epsilon+1} - \boldsymbol{x}^{\hat{k}_\epsilon} \parallel}{\alpha_{\hat{k}_\epsilon+1}} > \epsilon^\prime.
\end{equation}
Now, using similar analysis to that in \eqref{eq: |x_k -x_k-1| /alfa_k > eps}--\eqref{eq: f^k -f^k-1 < ...}, it is easy to show that
\begin{align}
F \left( \boldsymbol{x}^{\hat{k}_\epsilon+1} \right) - F \left( \boldsymbol{x}^{\hat{k}_\epsilon} \right)& \leq - \left( \dfrac{1}{\alpha_{\hat{k}_\epsilon+1}} ( 1- \dfrac{\parallel \nabla F \left( \boldsymbol{x}^{\hat{k}_\epsilon} \right)  - \boldsymbol{h}^{\hat{k}_\epsilon+1} \parallel}{\epsilon^\prime} ) - \frac{L_{\nabla f}}{2}   \right)  \parallel \boldsymbol{x}^{\hat{k}_\epsilon+1} - \boldsymbol{x}^{\hat{k}_\epsilon} \parallel^2.
\end{align}
Moreover, 
 for the considered sufficiently large $ \hat{k}_\epsilon $, there exist some $ \lambda_{\epsilon^\prime} >0 $ such that 
\begin{align}
  \dfrac{1}{\alpha_{\hat{k}_\epsilon+1}} ( 1- \dfrac{\parallel \nabla F \left( \boldsymbol{x}^{\hat{k}_\epsilon} \right)  - \boldsymbol{h}^{\hat{k}_\epsilon+1} \parallel}{\epsilon^\prime} ) - \frac{L_{\nabla F}}{2}   \geq \dfrac{\lambda_{\epsilon^\prime}}{\alpha_{\hat{k}_\epsilon+1}}.
\end{align}
Therefore, combining the above three inequalities results that
\begin{equation}
F \left( \boldsymbol{x}^{\hat{k}_\epsilon+1} \right) - F \left( \boldsymbol{x}^{\hat{k}_\epsilon} \right) \leq - {\epsilon^\prime}^2 \lambda_{\epsilon^\prime} \alpha_{\hat{k}_\epsilon+1} ,
\end{equation}
and hence, $ F \left( \boldsymbol{x}^{\hat{k}_\epsilon+1} \right) < F \left( \boldsymbol{x}^{\hat{k}_\epsilon} \right) $. Consequently, it follows that
\begin{align}
F \left( \boldsymbol{x}^{\hat{k}_\epsilon+1} \right) - F^\ast <  F \left( \boldsymbol{x}^{\hat{k}_\epsilon} \right) - F^\ast < \epsilon,
\end{align}
which indicates that $ \boldsymbol{x}^{\hat{k}_\epsilon+1}  \in \mathcal{X}_\epsilon $. Obviously, this contradicts  our initial assumption that $ \boldsymbol{x}^{\hat{k}_\epsilon+1}  \notin \bar{\mathcal{X}_\epsilon} $. Therefore, this assumption must be wrong and  $ \boldsymbol{x}^{\hat{k}_\epsilon+1}  \in \mathcal{X}_\epsilon $. 

The above result shows that under our proposed algorithm, once we enter the region $ \mathcal{X}_\epsilon $ (that we have shown in \eqref{eq: x_hat_k  in  X_eps} that it happens at a sufficiently large iteration $ \hat{k}_\epsilon $), we never go out of it. 
Consequently, for any $ \epsilon > 0 $, there exists a sufficiently large $ \hat{k}_\epsilon $ such that 
\begin{equation}
\forall k \geq  \hat{k}_\epsilon: \boldsymbol{x}^k \in \mathcal{X}_\epsilon,
\end{equation} 
or equivalently, 
\begin{equation}
\forall k \geq  \hat{k}_\epsilon: \quad  F \left( \boldsymbol{x}^k \right) - F ^\ast \leq \epsilon.
\end{equation}
Since  the above inequality is proved for any arbitrary $ \epsilon >0 $, it is concluded that 
\begin{equation}
\displaystyle\lim_{k \rightarrow \infty} F \left( \boldsymbol{x}^k \right) - F \left( \boldsymbol{x}^\ast \right) =0, 
\end{equation}
which completes the proof of Theorem \ref{th: conv_convex}. $ \hfill \blacksquare $

\subsection{Proof of Theorem \ref{th: conv2_nonconvex} (Non-convex case)}

Let $ \bar{X} $ be the set of limiting points of the proposed Algorithm \ref{alg: main}. Note that according to Lemma \ref{lem: x^k - x^k-1}, this set is not empty. In the following, we  prove that if a strictly local minimum of the optimization problem \eqref{eq: main stochastic optimisation} belongs to this set, then this set has only a single element
, i.e., the algorithm converges to the strictly local minimum point.

Let $ \bar{\boldsymbol{x}}^\ast $ denote the strictly local minimum point belonging to the set of limiting points $ \bar{X} $. Since $ \bar{\boldsymbol{x}}^\ast \in \bar{X} $, there exists a subsequence $  \left\lbrace \bar{k}_j \right\rbrace $ of the iteration indices 
so that the sequence 
$ \left\lbrace \boldsymbol{x}^{\bar{k}_j} \right\rbrace $ of the iterates generated by Algorithm \ref{alg: main} converges to $ \bar{\boldsymbol{x}}^\ast $, i.e.,
\begin{align}\label{eq: lim x_bar_k_j to bar_x_ast}
\displaystyle \lim_{j \rightarrow \infty} \boldsymbol{x}^{\bar{k}_j} = \bar{\boldsymbol{x}}^\ast.
\end{align}

For any $ r >0 $, let $ B\left( \bar{\boldsymbol{x}}^\ast, r \right) $ denote a ball of radius $ \epsilon>0 $ centred at $ \bar{\boldsymbol{x}}^\ast $. Now, 
for any sufficiently small $ \epsilon > 0 $ such that $ B\left( \bar{\boldsymbol{x}}^\ast,\epsilon \right) $ contains only one stationary point (such a non-empty region exists because $ \bar{\boldsymbol{x}}^\ast $ is 
a strictly local minimum point),  we define the following sets of points:
\begin{equation}\label{eq: def of X_eps_tilde}
\widetilde{\mathcal{X}}_\epsilon \triangleq \left\lbrace  \boldsymbol{x}  \in \mathcal{X} \left| F \left( \boldsymbol{x} \right) - F \left( \bar{\boldsymbol{x}}^\ast \right)  < \epsilon \right. \right\rbrace \cap B\left(\bar{\boldsymbol{x}}^\ast, \epsilon \right)  ,
\end{equation}
\begin{equation}\label{eq: def of bar_X_eps_tilde}
\bar{\widetilde{\mathcal{X}}_\epsilon}  \triangleq \left( \mathcal{X} - \widetilde{\mathcal{X}}_\epsilon \right) \cap B\left( \bar{\boldsymbol{x}}^\ast,\epsilon \right) , 
\end{equation}
and the set of indices
\begin{equation}\label{eq: def of bar_K_eps_tilde}
\bar{\widetilde{\mathcal{K}_\epsilon}}  \triangleq \left\lbrace  k \left| \boldsymbol{x}^k  \in \bar{\widetilde{\mathcal{X}}_\epsilon} \right. \right\rbrace.
\end{equation}

In the following, we will show that after a sufficiently large iteration, 
we will not leave the region $  \widetilde{\mathcal{X}}_\epsilon$. Therefore, as this result is proved for any arbitrarily small $ \epsilon>0 $, the whole Algorithm \ref{alg: main} converges to the strictly local minimum $ \bar{\boldsymbol{x}}^\ast $. For this purpose, first note that following from \eqref{eq: lim x_bar_k_j to bar_x_ast}, 
there exists some sufficiently large $ \bar{k}_\epsilon \in  \left\lbrace \bar{k}_j \right\rbrace $ such that 
\begin{equation}\label{eq: x_k_bar_eps in tilde_X_eps}
\boldsymbol{x}^{\bar{k}_\epsilon} \in \widetilde{\mathcal{X}}_{\frac{\epsilon}{2}},
\end{equation} 
and hence $ \boldsymbol{x}^{\bar{k}_\epsilon} \in \widetilde{\mathcal{X}}_\epsilon $. 
We also prove by contradiction that  $ \boldsymbol{x}^{\bar{k}+1}  \in \widetilde{\mathcal{X}}_\epsilon $. 
Suppose that $ \boldsymbol{x}^{\bar{k}_\epsilon+1}  \notin \widetilde{\mathcal{X}}_\epsilon $, then 
one of the following two cases arises:

\textbf{Case 1:} If $ \boldsymbol{x}^{\bar{k}_\epsilon+1} \notin B\left( \bar{\boldsymbol{x}}^\ast,\epsilon \right)  $,  using the fact that  $ \boldsymbol{x}^{\bar{k}_\epsilon} \in B\left( \bar{\boldsymbol{x}}^\ast,\epsilon/2 \right)  $, it follows that
\begin{equation}\label{eq: x_k_bar+1 -x_k_bar > eps/2}
\parallel \boldsymbol{x}^{\bar{k}_\epsilon+1} - \boldsymbol{x}^{\bar{k}_\epsilon} \parallel > \dfrac{\epsilon}{2},
\end{equation}
and hence
\begin{align}
\dfrac{\parallel \boldsymbol{x}^{\bar{k}_\epsilon+1} - \boldsymbol{x}^{\bar{k}_\epsilon} \parallel}{\alpha_{\bar{k}_\epsilon}} &> \dfrac{\epsilon}{2\alpha_{\bar{k}_\epsilon}}  > \epsilon,
\end{align}
where the last inequality follows from the fact that $ \alpha_k \rightarrow 0 $. Using the above inequality and following similar analysis to that in \eqref{eq: |x_k -x_k-1| /alfa_k > eps}--\eqref{eq: f^k -f^k-1 < ...}, it is easily shown that
\begin{align}
F \left( \boldsymbol{x}^{\bar{k}_\epsilon+1} \right) - F \left( \boldsymbol{x}^{\bar{k}_\epsilon} \right)& \leq - \left( \dfrac{1}{\alpha_{\bar{k}_\epsilon+1}} ( 1- \dfrac{\parallel \nabla F \left( \boldsymbol{x}^{\bar{k}_\epsilon} \right)  - \boldsymbol{h}^{\bar{k}_\epsilon+1} \parallel}{\epsilon} ) - \frac{L_{\nabla F}}{2}   \right)  \parallel \boldsymbol{x}^{\bar{k}_\epsilon+1} - \boldsymbol{x}^{\bar{k}_\epsilon} \parallel^2, \notag \\
& < - \left( \dfrac{1}{\alpha_{\bar{k}_\epsilon+1}} ( 1- \dfrac{\parallel \nabla F \left( \boldsymbol{x}^{\bar{k}_\epsilon} \right)  - \boldsymbol{h}^{\bar{k}_\epsilon+1} \parallel}{\epsilon} ) - \frac{L_{\nabla F}}{2}   \right) \dfrac{\epsilon}{2},
\end{align}
where the last inequality directly follows from \eqref{eq: x_k_bar+1 -x_k_bar > eps/2}. 
Now, considering the fact that $ \left\lbrace \alpha_k \right\rbrace $ is a monotonically decreasing sequence converging to $ 0 $, for sufficiently large $ \bar{k}_\epsilon $, we will have
\begin{equation}
F \left( \boldsymbol{x}^{\bar{k}_\epsilon+1} \right) - F \left( \boldsymbol{x}^{\bar{k}_\epsilon} \right) < -2c,
\end{equation}
where $ c $ is the maximum value of $ | F (.) | $ over the region $ \mathcal{X} $. This, obviously, contradicts with the fact that $ c < \infty $ (derived from Assumption \ref{assump: stoch opt formulation}). Therefore, the statement of Case 1 is false and this case can not arise.

\textbf{Case 2:} If $ \boldsymbol{x}^{\bar{k}_\epsilon+1} \in B\left( \bar{\boldsymbol{x}}^\ast,\epsilon \right)  $, then the following lemma, which is similar to Lemma \ref{lem: convex, k in bar_K_eps}, holds.

\begin{lemma}
\label{lem: nonconvex, k in bar_K_eps}
Under the conditions of Theorem \ref{th: conv2_nonconvex} and the definitions in \eqref{eq: def of X_eps_tilde}--\eqref{eq: def of bar_K_eps_tilde}, 
there exists a $ \epsilon^{''} > 0 $ such that under Algorithm \ref{alg: main}, 
we have
\begin{equation}\label{eq: >eps_prime}
\dfrac{\parallel \boldsymbol{x}^k - \boldsymbol{x}^{k-1} \parallel}{\alpha_k} > \epsilon^{''}, \quad  \forall k \in \bar{\widetilde{\mathcal{K}_\epsilon}} .
\end{equation}
\end{lemma}

\proof   ~
 First note that if $ | \bar{\widetilde{\mathcal{K}_\epsilon}}  | < \infty $, then the statement of the lemma is obvious. Therefore, it suffices to prove the lemma for the case where $ | \bar{\widetilde{\mathcal{K}_\epsilon}}  | = \infty $. We prove this case by contradiction, as follows. 
Suppose that there exists no $ \epsilon^{''} >0 $ that satisfies \eqref{eq: >eps_prime}. Therefore, there should exist a  subsequence 
$ \left\lbrace k^{''}_j \right\rbrace \subset \bar{\widetilde{\mathcal{K}_\epsilon}}  $ such that  
\begin{equation}\label{eq: lim x''- x'' /alfa'' = 0}
\lim_{j \rightarrow \infty}  \dfrac{\parallel \boldsymbol{x}^{k^{''}_j} - \boldsymbol{x}^{k^{''}_j-1} \parallel}{\alpha_{k^{''}_j} } =0.
\end{equation}
This means that the sequence $ \left\lbrace \boldsymbol{x}_{k^{''}_j} \right\rbrace $ is convergent to some limit point. Let $ \tilde{\boldsymbol{x}}=\left( \tilde{\boldsymbol{x}}_l \right)_{l=1}^I $ be the limit point of this convergent subsequence. 
Note that since $ \left\lbrace k^{''}_j \right\rbrace \subset \bar{\widetilde{\mathcal{K}_\epsilon}}  $ and due to the definition of $ \bar{\widetilde{\mathcal{K}_\epsilon}} $ in \eqref{eq: def of bar_K_eps_tilde}, it follows that $ \left\lbrace \boldsymbol{x}_{k^{''}_j} \right\rbrace \in B\left( \bar{\boldsymbol{x}}^\ast,\epsilon \right) $, and hence, since $ B\left( \bar{\boldsymbol{x}}^\ast,\epsilon \right) $ is a closed and convex set, the  limit point $ \tilde{\boldsymbol{x}}$ belongs to $ B\left( \bar{\boldsymbol{x}}^\ast,\epsilon \right) $, as well. 
Using similar analysis to that in \eqref{eq: first-order opt cond_v3} for the convergent sequence $ \left\lbrace \boldsymbol{x}_{k^{''}_j} \right\rbrace $ along with \eqref{eq: lim x''- x'' /alfa'' = 0}, 
it is easy to verify that
\begin{equation}
\langle \boldsymbol{x} -  \tilde{\boldsymbol{x}} , \nabla f \left( \tilde{\boldsymbol{x}} \right)  \rangle \geq 0, \quad \forall \boldsymbol{x} \in \mathcal{X},
\end{equation}
and hence the limit point $ \tilde{\boldsymbol{x}} $ is a stationary point to the optimization problem \eqref{eq: main stochastic optimisation}. 
Considering  $ \tilde{\boldsymbol{x}} \in B\left( \bar{\boldsymbol{x}}^\ast,\epsilon \right) $ and the fact that there is only one stationary point (i.e., $ \bar{\boldsymbol{x}}^\ast  $) belonging to $ B\left( \bar{\boldsymbol{x}}^\ast,\epsilon \right) $, it is concluded that $ \tilde{\boldsymbol{x}} = \bar{\boldsymbol{x}}^\ast  $, and hence we have $ F \left( \tilde{\boldsymbol{x}} \right) = F \left( \bar{\boldsymbol{x}}^\ast \right)  $, or equivalently, 
\begin{equation}
\displaystyle \lim_{j \rightarrow \infty}  F \left( \boldsymbol{x}^{k^{''}_j} \right) = F \left( \bar{\boldsymbol{x}}^\ast \right).
\end{equation} 
Accordingly, for any fixed $ \gamma > 0 $, there exists some sufficiently large $ k^{''}_\gamma $ such that
\begin{equation}
\forall k^{''}_j \geq k^{''}_\gamma  : \quad F \left( \boldsymbol{x}^{k^{''}_j} \right) - F \left( \bar{\boldsymbol{x}}^\ast \right) < \gamma.
\end{equation}
Substituting $ \gamma = \epsilon $, it follows that 
for sufficiently large $ k^{''}_j \in \bar{\widetilde{\mathcal{K}_\epsilon}}  $, we have $ F \left( \boldsymbol{x}^{k^{''}_j} \right) - F \left( \bar{\boldsymbol{x}}^\ast \right) < \epsilon $. This means that $ \boldsymbol{x}^{k^\prime_j} \in \widetilde{\mathcal{X}}_\epsilon $, which is obviously a contradiction of the fact that $ k^{''}_j \in \bar{\widetilde{\mathcal{K}_\epsilon}} $. 
Therefore, the contradiction assumption must be wrong, and this completes the proof of the lemma. 
 $ \hfill \blacksquare $

In the rest of this subsection, we use the above lemma to prove that under Case 2, Algorithm \ref{alg: main} will converge to the strictly local minimum $ \bar{\boldsymbol{x}}^\ast $. Recall that for the sufficiently large $ \bar{k}_\epsilon $, we have $  \boldsymbol{x}^{\bar{k}_\epsilon} \in \widetilde{\mathcal{X}}_\epsilon  $
and $ \boldsymbol{x}^{\bar{k}_\epsilon+1} \in B\left( \bar{\boldsymbol{x}}^\ast,\epsilon \right)  $ (as considered in Case 2). In the following, we will show by contradiction that $  \boldsymbol{x}^{\bar{k}_\epsilon+1} \in \widetilde{\mathcal{X}}_\epsilon  $, as well. 
%
%
For this purpose, first suppose $ \boldsymbol{x}^{\bar{k}_\epsilon+1}  \notin \widetilde{\mathcal{X}}_\epsilon $, and hence, $ \boldsymbol{x}^{\bar{k}_\epsilon+1}  \in \bar{\widetilde{\mathcal{X}}_\epsilon} $. Therefore, according to Lemma \ref{lem: nonconvex, k in bar_K_eps}, there exists a fixed  $ \epsilon^{''} > 0 $ such that
\begin{equation}
\dfrac{ \parallel \boldsymbol{x}^{\bar{k}_\epsilon+1} - \boldsymbol{x}^{\bar{k}_\epsilon} \parallel}{\alpha_{\bar{k}_\epsilon+1}} > \epsilon^{''} .
\end{equation}
Now, using similar analysis to that in \eqref{eq: |x_k -x_k-1| /alfa_k > eps}--\eqref{eq: f^k -f^k-1 < ...}, it is easy to show that
\begin{align}
F \left( \boldsymbol{x}^{\bar{k}_\epsilon+1} \right) - F \left( \boldsymbol{x}^{\bar{k}_\epsilon} \right)& \leq - \left( \dfrac{1}{\alpha_{\bar{k}_\epsilon+1}} ( 1- \dfrac{\parallel \nabla F \left( \boldsymbol{x}^{\bar{k}_\epsilon} \right)  - \boldsymbol{h}^{\bar{k}_\epsilon+1} \parallel}{\epsilon^{''} } ) - \frac{L_{\nabla F}}{2}   \right)  \parallel \boldsymbol{x}^{\bar{k}_\epsilon+1} - \boldsymbol{x}^{\bar{k}_\epsilon} \parallel^2.
\end{align}
Moreover, 
 for the considered sufficiently large $ \bar{k}_\epsilon $ there exist some $ \lambda_{\epsilon^{''} } >0 $ such that 
\begin{align}
  \dfrac{1}{\alpha_{\bar{k}_\epsilon+1}} ( 1- \dfrac{\parallel \nabla F \left( \boldsymbol{x}^{\bar{k}_\epsilon} \right)  - \boldsymbol{h}^{\bar{k}_\epsilon+1} \parallel}{\epsilon^{''} } ) - \frac{L_{\nabla F}}{2}   \geq \dfrac{\lambda_{\epsilon^{''} }}{\alpha_{\bar{k}_\epsilon+1}}.
\end{align}
Therefore, combining the above three inequalities results that
\begin{equation}
F \left( \boldsymbol{x}^{\bar{k}_\epsilon+1} \right) - F \left( \boldsymbol{x}^{\bar{k}_\epsilon} \right) \leq - {\epsilon^{''}}^2 \lambda_{\epsilon^{''}} \alpha_{\bar{k}_\epsilon+1} ,
\end{equation}
and hence, $ F \left( \boldsymbol{x}^{\bar{k}_\epsilon+1} \right) < F \left( \boldsymbol{x}^{\bar{k}_\epsilon} \right) $. Consequently, it follows that
\begin{align}
F \left( \boldsymbol{x}^{\bar{k}_\epsilon+1} \right) - F^\ast <  F \left( \boldsymbol{x}^{\bar{k}_\epsilon} \right) - F^\ast < \epsilon,
\end{align}
which along with the fact that $ \boldsymbol{x}^{\bar{k}_\epsilon+1} \in B\left( \bar{\boldsymbol{x}}^\ast,\epsilon \right)  $ (as considered in Case 2) indicates that $ \boldsymbol{x}^{\bar{k}_\epsilon+1}  \in \widetilde{\mathcal{X}}_\epsilon $. Obviously, this contradicts  our 
contradiction assumption 
 that $ \boldsymbol{x}^{\bar{k}_\epsilon+1}  \notin \widetilde{\bar{\mathcal{X}}_\epsilon} $. Therefore, this assumption must be wrong and $ \boldsymbol{x}^{\bar{k}_\epsilon+1}  \in \widetilde{\mathcal{X}}_\epsilon $. 

The above result shows that for the conditions in Theorem \ref{th: conv2_nonconvex}, under our proposed algorithm, once we enter the region $ \widetilde{\mathcal{X}}_\epsilon $ (that we have shown in \eqref{eq: x_k_bar_eps in tilde_X_eps} that it happens at some sufficiently large iteration $ \bar{k}_\epsilon $), we never go out of it. 
Consequently, for any $ \epsilon > 0 $, there exists a sufficiently large $ \bar{k}_\epsilon $ such that 
\begin{equation}
\forall k \geq  \bar{k}_\epsilon: \boldsymbol{x}^k \in \widetilde{\mathcal{X}}_\epsilon.
\end{equation} 
Hence,
using the definition of $ \widetilde{\mathcal{X}}_\epsilon $ in \eqref{eq: def of X_eps_tilde}, it follows that 
\begin{equation}
\forall k \geq  \bar{k}_\epsilon: \quad  F \left( \boldsymbol{x}^k \right) - F \left( \bar{\boldsymbol{x}}^\ast \right)  \leq \epsilon.
\end{equation} 
Since this result is proved for any arbitrarily small $ \epsilon >0 $, it is concluded that 
\begin{equation}
\displaystyle\lim_{k \rightarrow \infty} F \left( \boldsymbol{x}^k \right) - F \left( \bar{\boldsymbol{x}}^\ast \right)  =0, 
\end{equation}
which completes the proof of Theorem \ref{th: conv2_nonconvex}. $ \hfill \blacksquare $

\end{document}